\documentclass[aps,pra,reprint,10pt,twocolumn,superscriptaddress,floatfix]{revtex4-1}

\usepackage{times}
\usepackage{amsmath,amssymb,bm,graphicx}
\usepackage[breaklinks,colorlinks=true,urlcolor=blue,citecolor=blue,linkcolor=blue]{hyperref}
\usepackage{mathtools}
\usepackage{xcolor}
\usepackage{enumitem}
\usepackage{braket}
\usepackage{bbm}

\newcommand{\bigext}{jpg}

\def\rvec{\mathbf{r}}
\def\avec{\mathbf{a}}
\def\kvec{\mathbf{k}}

\begin{document}

\title{Majorana lattices from the quantized Hall limit of a proximitized spin-orbit coupled electron gas}

\date{\today}

\author{Ryan V. Mishmash}
\affiliation{Department of Physics, Princeton University, Princeton, New Jersey 08540, USA}
\affiliation{Department of Physics, University of California, Berkeley, California 94720, USA}

\author{A. Yazdani}
\affiliation{Department of Physics, Princeton University, Princeton, New Jersey 08540, USA}

\author{Michael P. Zaletel}
\affiliation{Department of Physics, Princeton University, Princeton, New Jersey 08540, USA}
\affiliation{Department of Physics, University of California, Berkeley, California 94720, USA}

\begin{abstract}
Motivated by recent experiments demonstrating intricate quantum Hall physics on the surface of elemental bismuth, we consider 
proximity coupling an $s$-wave superconductor to a two-dimensional electron gas with strong Rashba spin-orbit interactions in the presence of a strong perpendicular magnetic field. We focus on the high-field limit so that the superconductivity can be treated as a perturbation to the low-lying Landau levels. In the clean case, wherein the superconducting order parameter takes the form of an Abrikosov vortex lattice, we show that a lattice of hybridized Majorana modes emerges near the plateau transition of the lowest Landau level. However, unless magnetic-symmetry-violating perturbations are present, the system always has an even number of chiral Majorana edge modes and thus is strictly speaking Abelian in nature, in agreement with previous work on related setups. Interestingly, however, a \emph{weak} topological superconducting phase can very naturally be stabilized near the plateau transition for the square vortex lattice. The relevance of our findings to potential near-term experiments on proximitized materials such as bismuth will be discussed.
\end{abstract}

\maketitle

\section{Introduction} \label{sec:intro}

The past decade has witnessed a flurry of activity, both theoretical and experimental alike, on engineering topological phases of matter by piecing together less exotic, more well-understood components. For example, a common paradigm involves realizing spinless $p$-wave superconductivity, in either one \cite{Kitaev01_PhysU_44_131} or two \cite{read_paired_2000} dimensions, by proximity coupling (much more abundant) $s$-wave superconductors with spin-orbit coupled materials in the presence of modest magnetic (Zeeman) fields \cite{lutchyn_majorana_2010,oreg_helical_2010,sau_generic_2010,alicea_majorana_2010,sau_non-abelian_2010} \footnote{A closely related approach involves proximity coupling an $s$-wave superconductor to the edge of a 2D or 3D topological insulator \cite{fu_superconducting_2008,fu_josephson_2009}.}. The one-dimensional (1D) version of this pursuit has achieved remarkable experimental maturity in the past few years; and while still hotly debated, the effort has led to mounting evidence for the existence of Majorana modes at the ends of 1D superconductor-semiconductor heterostructure devices (see Ref.~\cite{lutchyn_majorana_2018} for a recent review) and ferromagnetic atom chains on the surface of strongly spin-orbit coupled superconductors \cite{nadj-perge_observation_2014}.

In the abovementioned proposals, 
the primary role of the magnetic field is to Zeeman split the bands at zero momentum, thereby producing a ``one-band'' regime across which $s$-wave pairing can, due to presence of the spin-orbit coupling, pair states on opposite sides of the Fermi surface \cite{alicea_new_2012}. Here, we focus on the case of two spatial dimensions (2D) in a quite different physical regime. We consider applying a strong perpendicular magnetic field to an $s$-wave superconductor proximity coupled to a Rashba spin-orbit coupled two-dimensional electron gas (2DEG). In this situation, orbital effects due to the magnetic field dominate: screening of the magnetic field by the superconductor draws the field into $hc/2e$ flux tubes while preserving the average flux, which subsequently organizes the single-particle states of the 2DEG into Landau levels (on the other hand, Zeeman effects are likely less important).

We consider a minimal Hamiltonian for this heterostructure setup,
\begin{align}
H &= H_\mathrm{2DEG} + H_\Delta, \label{eq:H}
\end{align}
where
\begin{align}
H_\mathrm{2DEG} &= \int d^2 r \, \Psi^\dagger \biggl[\frac{(\mathbf{p} - \frac{e}{c} \mathbf{A})^2}{2 m} - \mu + V(\mathbf{r}) \label{eq:H2DEG} \\
& ~~~~~~~~~~~~~~~ - \alpha_R \, \bm{\sigma} \times \left(\mathbf{p} - \frac{e}{c} \mathbf{A}\right) \cdot \hat{z} + E_Z \sigma^z \biggr] \Psi, \nonumber \\
H_\Delta &= \int d^2 r \, \Psi^\dagger_\uparrow\,\Delta(\mathbf{r}) \Psi^\dagger_\downarrow + \mathrm{H.c.} \label{eq:HDelta}
\end{align}
Here $e$ is the (single) electron charge; $m$ is the effective electron mass; $\mu$ is the chemical potential; $\alpha_R$ and $E_Z$ are the Rashba and Zeeman coupling strengths, respectively; $\mathbf{A} = \mathbf{A}(\mathbf{r})$ is the vector potential (with $\mathbf{B}(\mathbf{r}) = \nabla \times \mathbf{A}(\mathbf{r}) = B(\mathbf{r})\hat{z}$ the magnetic field felt by the electrons); $V(\mathbf{r})$ is a scalar potential; $\Delta(\mathbf{r})$ is the superconducting pair field; and the $\sigma^j$ are Pauli matrices acting in the spin space $\Psi^\dagger=(\Psi_\uparrow^\dagger, \Psi_\downarrow^\dagger)$. $\mathbf{B}(\mathbf{r})$, $V(\mathbf{r})$, and $\Delta(\mathbf{r})$ will in general all be spatially nonuniform due to formation of vortices in the superconductor.

We assume the high-field limit of this problem, so that the cyclotron gap $\hbar \omega_c$ is the largest energy scale, and focus primarily on the case of a clean system. Additionally, we take the external magnetic field to be near the upper critical field $H_{c2}$ of the assumed type-II superconductor. This allows us to (\emph{i}) use for $\Delta(\rvec)$ the well-known lowest-Landau-level form for the Abrikosov vortex lattice arising in Ginzburg-Landau theory \cite{michael_tinkham_introduction_1996,rosenstein_ginzburg-landau_2010} (see Fig.~\ref{fig:lattices}) and (\emph{ii}) ignore at leading order screening-induced inhomogeneities in the magnetic field, i.e., $B(\rvec) \approx B \hat{z}$. In contrast to most previous approaches \cite{liu_electronic_2015,murray_majorana_2015,ariad_how_2018}, we are then able to employ a standard Landau gauge and write $H_\mathrm{\Delta}$ in the corresponding magnetic Bloch basis appropriate for the vortex lattice solution, considering both square and triangular vortex lattices simultaneously. (An exception is Ref.~\cite{zocher_topological_2016}, where the authors take a similar approach in the related context of a spinless $p+ip$ superconductor in the presence of a triangular vortex lattice.) Furthermore, assuming that the pairing strength is sufficiently small relative to the Landau-level spacing, we can project the Bogoliubov-de Gennes (BdG) Hamiltonian into the lowest Landau level (LLL) of $H_\mathrm{2DEG}$. The resulting BdG problem closely resembles that of a spinless $p+ip$ superconductor \cite{read_paired_2000}, albeit with a modified odd-parity gap function---denoted $\tilde\Delta_{10}(\kvec)$ in Sec.~\ref{sec:RashbaPD} below---which exhibits an intricate structure with multiple Dirac nodes. That is, purely orbital effects and spin-orbit coupling conspire to provide an intriguing means of obtaining spinless superconductivity deep in the quantum Hall regime.

Sweeping the chemical potential $\mu$ through the LLL gives rise to a quantum Hall plateau transition, the nature of which is governed by the pairing $\tilde\Delta_{10}(\kvec)$ rather than disorder. If the magnetic translation symmetry of the vortex lattice is preserved, we find that all phases in the vicinity of the plateau transition are necessarily Abelian with integer Chern number $C$ (in a convention where a single copy of the integer quantum Hall effect has $C=1$).  Preclusion of non-Abelian half-integer $C$ states due to magnetic symmetry was also pointed out recently in a related context by Jeon et al.~\cite{jeon_topological_2018} and also earlier in Ref.~\cite{zocher_topological_2016}. Interestingly, in the case of a square vortex lattice we find that the intermediate-$\mu$ phases are \emph{weak} topological superconductors,
and consequently a dislocation in the vortex lattice will trap an unpaired Majorana zero mode. 

All of the realized Abelian integer-Chern states harbor a lattice of hybridized Majorana modes (located at the positions of the original vortices)---a highly nontrivial physical system which has garnered significant recent theoretical attention \cite{grosfeld_electronic_2006,ludwig_two-dimensional_2011,lahtinen_topological_2012,kraus_majorana_2011,laumann_disorder-induced_2012,zhou_hierarchical_2013,zhou_topological_2014,biswas_majorana_2013,silaev_majorana_2013,liu_electronic_2015,murray_majorana_2015,ariad_how_2018}. By taking the approach summarized above, we are able to provide both a fresh theoretical perspective on this system, as well as a straightforward recipe for how to realize such a Majorana lattice in experiment. Importantly, by working in the high-field limit where we can project into a single Landau level, the \emph{entirety} of the low-energy density of states necessarily corresponds to a band of Majorana modes: a Landau level has one fermionic mode (quantum dimension 2) per flux quantum $hc/e$, while each vortex carries one \emph{superconducting} flux quantum $hc/2e$; hence, the number of states per vortex is $\frac12$ (a Majorana, with quantum dimension $\sqrt{2}$).
All states other than the Majorana modes are separated away by an energy on the order of $\hbar \omega_c$, which can easily be $\sim$10\,meV. This is in contrast to other platforms and regimes wherein the density of states is polluted by other low-energy modes in the vortex cores \cite{sun_majorana_2016,wang_evidence_2018,liu_robust_2018} \footnote{On the other hand, topological defects in the spin-orbit coupling \cite{sato_non-abelian_2009,sau_non-abelian_2010}, as opposed to in the phase of the superconducting order parameter, were reported in a recent experiment \cite{menard_isolated_2018} to harbor robust, isolated Majorana modes in the absence of such subgap states.}.

We now discuss potential laboratory realizations of the above system and associated considerations that arise. Indeed, models such as Eq.~\eqref{eq:H} are now on the near-term experimental horizon in light of recent advances in epitaxial growth for superconductor-semiconductor hybrid devices (see, e.g., Ref.~\cite{krogstrup_epitaxy_2015}). Proximitizing the strongly Rashba spin-orbit coupled surface states of elemental bismuth \cite{koroteev_strong_2004,hofmann_surfaces_2006} is one enticing possibility. Bismuth is an extremely clean electronic system (with a bulk mean free path on the order of mm) which has long played a central role in the development of electronic characterization techniques \cite{edelman_electrons_1976} and has more recently become a playground for novel topological phenomena \cite{drozdov_one-dimensional_2014,li_magnetic_2014,murani_ballistic_2017,murakami_quantum_2006,reis_bismuthene_2017,schindler_higher-order_2018}. Recently, by performing scanning tunneling microscope (STM) measurements on the (unproximitized) Bi(111) surface in high perpendicular magnetic fields, Feldman et al.~\cite{feldman_observation_2016} provided evidence for the emergence of a gapped nematic quantum Hall state (arising from a combination of local strain and Coulomb interactions spontaneously lifting the six-fold Landau-level valley degeneracy characteristic of the anisotropic hole states on this surface). However, any such Rashba coupled surface could in principle suffice for the 2DEG portion of the system in our setup, so as long as it can be grown epitaxially with a sizable proximity effect on the surface of a strong type-II superconductor. We note that while our simplified model, Eq.~\eqref{eq:H}, is seemingly far-removed from complicated band structures like the Bi(111) surface, it can be viewed as a (fully isotropic) proxy for Landau-level states arising from a single electron pocket centered about the $\Gamma$ point in a real material.

As discussed above, we approach Eq.~\eqref{eq:H} from the rather unusual limit in which the proximitized superconductivity can be treated as a perturbation to the Landau levels, an assumption which amounts to working in the regime where the cyclotron gap is much larger than the characteristic $s$-wave proximity-induced pairing gap $\Delta_0$: $\hbar\omega_c = \frac{\hbar eB}{mc} \gg \Delta_0$. To obtain this limit, it is thus of course desirable to use a 2DEG whose carriers have a small effective mass, as is the case for Bi(111) \cite{hofmann_surfaces_2006,feldman_observation_2016}. Furthermore, for our analysis to apply, we require $\Delta_0$ to be much larger than the characteristic disorder and electron-electron interaction strengths (we briefly address the former in Sec.~\ref{sec:disorder}; the latter certainly constitutes an intriguing issue which we leave for future work and comment on in our concluding remarks in Sec.~\ref{sec:discussion}). In a real experiment, this will require a superconductor with a sufficiently large $H_{c2}$ so as to be able to withstand the requisite large magnetic fields and still have appreciable $\Delta_0$. Promising candidate superconductors include epitaxially grown thin films of FeSe or NbN: both of which show upper critical fields of $\sim$16\,T, and in both cases, previous experiments have been able to image vortex lattices using spectroscopic mapping with the STM \cite{song_direct_2011,wang_atomically_2017}. The spin-orbit coupled 2DEG---such as Bi(111)---has to be grown epitaxially on the surface of these thin film superconductors at a thickness that is comparable or below the coherence length (e.g., 5\,nm for FeSe) to allow for at least a weak superconducting gap to develop on the Landau levels. Thus, in addition to simplifying the theoretical analysis in several respects, the high-field limit is also quite experimentally reasonable.

Finally, while our ultimate goal here is geared more towards engineering a Majorana lattice and not necessarily a non-Abelian topological superconductor with half-integer Chern number \cite{read_paired_2000}, the latter naturally arises near the plateau transition in our model provided that the magnetic translation symmetry is broken (for example by a unit-cell-doubling superlattice potential) \cite{zocher_topological_2016}. Even without such doubling, however, the resulting symmetry-respecting phases for the square vortex lattice are of interest for topological quantum computing applications \cite{nayak_non-abelian_2008} as dislocations in the vortex lattice trap an unpaired Majorana zero mode. We note that our work is closely related to previous work by Qi, Hughes, and Zhang \cite{qi_chiral_2010}, in which the authors discussed a similar setup, but focused mainly on proximity coupling a quantum anomalous Hall state. In contrast, we here consider the analogous problem of proximity coupling a (Rashba spin-orbit coupled) 2DEG in the quantum Hall regime, while also carefully accounting for the presence of vortices in the superconductor.

The rest of the paper is organized as follows. In Sec.~\ref{sec:setup}, we describe the technical details underlying our calculations, including reviews of the high-field Abrikosov vortex lattice solution (Sec.~\ref{sec:AVL}) and the Landau-level structure of a Rashba-coupled 2DEG (Sec.~\ref{sec:RashbaLL}); the most important part is Sec.~\ref{sec:MBB} wherein we derive expressions for the gap function in the appropriate magnetic Bloch basis. Appendix~\ref{app:details} contains additional details relevant to Sec.~\ref{sec:setup}. Secs.~\ref{sec:PD} and \ref{sec:weak} contain our main results. Sec.~\ref{sec:PD} concerns the nature of the quantum Hall plateau transition in the vicinity of the LLL. In Sec.~\ref{sec:spinful}, as a warmup, we first consider the lowest spinful Landau level in the absence of Rashba coupling. Then, in Sec.~\ref{sec:RashbaPD} we work out the much more interesting case of the lowest spinless Rashba-coupled Landau level discussed above. The weak topological phases harboring Majorana zero modes at lattice dislocations, which emerge near this plateau transition, are described in Sec.~\ref{sec:weak}. In Sec.~\ref{sec:disorder}, we briefly discuss the effects of disorder on the system, and we finally conclude in Sec.~\ref{sec:discussion}.

\section{Setup and technical ingredients} \label{sec:setup}

\subsection{Abrikosov vortex lattice in the high-field limit} \label{sec:AVL}

In the case of a clean system, the applied perpendicular magnetic field induces a perfect Abrikosov vortex lattice in the superconductor \cite{michael_tinkham_introduction_1996}. The spatial dependence of $B(\mathbf{r})$, $V(\mathbf{r})$, and $|\Delta(\mathbf{r})|$ in Eq.~\eqref{eq:H} will thus all have a periodicity given by the resulting Bravais lattice. For technical simplicity, we choose for $\Delta(\mathbf{r})$ the well-known solution of the Ginzburg-Landau (GL) equations valid \emph{at} the upper critical field $H_{c2}$. In this limit, which is not inconsistent with the high-field limit described in Sec.~\ref{sec:intro}, screening of the magnetic field by the superconductor can be ignored at leading order in $\Delta$ \cite{michael_tinkham_introduction_1996}; one then solves the ``linearized Ginzburg-Landau'' equation, i.e., the single-particle Schr\"{o}dinger equation for charge $2e$ particles in a constant background magnetic field $\mathbf{B}_0 = \nabla \times \mathbf{A}_0$ (equivalent to the applied external field). Working in a Landau gauge $\mathbf{A}_0 = B x \hat{y}$, the resulting lowest-Landau-level vortex lattice solution at $H_{c2}$ reads \cite{michael_tinkham_introduction_1996}
\begin{align}
\Delta(\mathbf{r}) = \sum_{j=-\infty}^\infty C_j \, e^{i k_j y - \frac{1}{2 \xi^2} (x - x_j)^2}, \label{eq:DLLL}
\end{align}
where $k_j \equiv \frac{2\pi j}{a}$ (with $a$ the intervortex separation), $x_j \equiv \frac12 k_j \ell_B^2$, and $\xi \equiv \frac{\ell_B}{\sqrt{2}}$ with $\ell_B = \sqrt{\hbar c / e B}$ the magnetic length for the charge $e$ electrons of $H_\mathrm{2DEG}$ \footnote{Note that Eq.~\eqref{eq:DLLL} is merely a linear superposition of LLL wave functions for charge $2e$ particles (or equivalently charge $e$ particles feeling twice the magnetic flux).}.

\begin{figure}[t]
\begin{center} 
\includegraphics[width=\columnwidth]{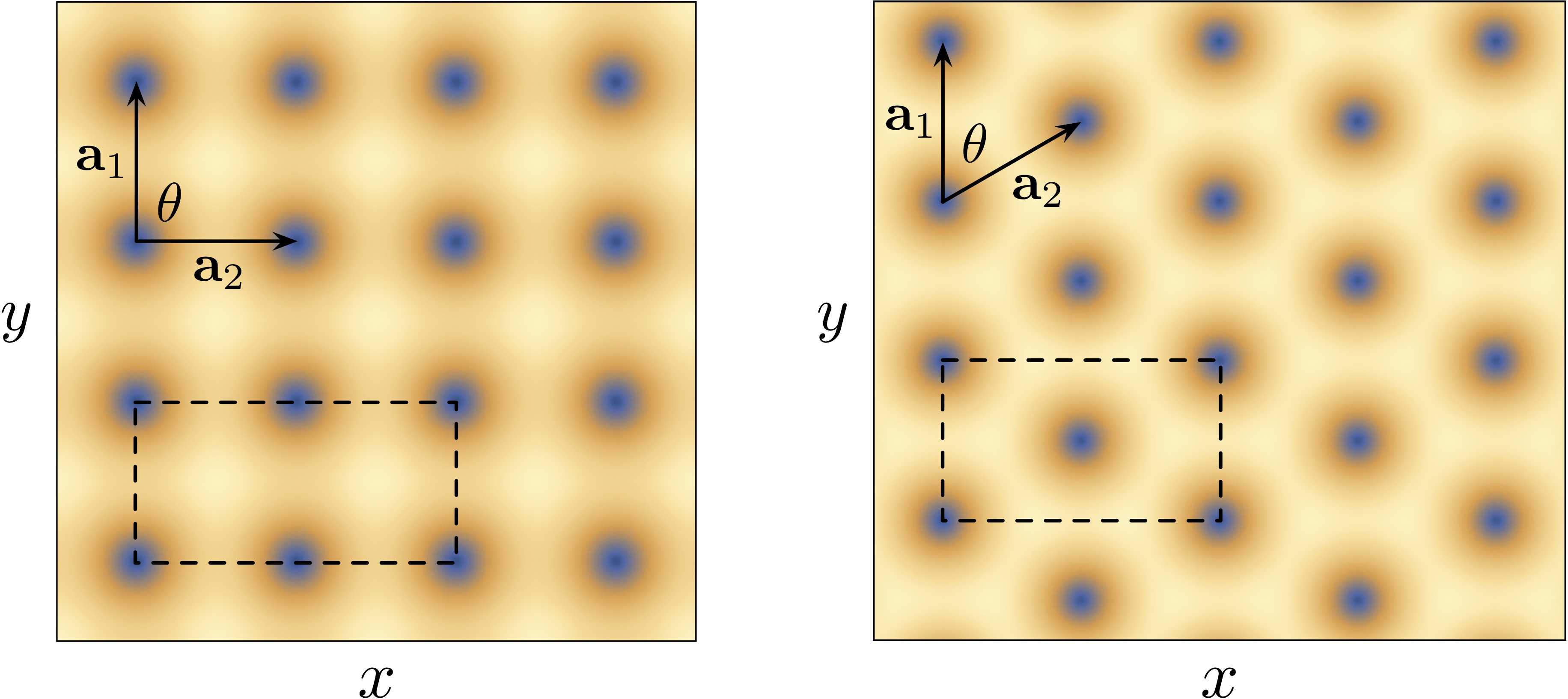}
\caption{Spatial dependence of the pairing potential, $|\Delta(\rvec)|$, for the square (left) and triangular (right) Abrikosov vortex lattice solutions [see Eqs.~\eqref{eq:DLLL} and \eqref{eq:Cj}]. Also shown are the respective primitive translation vectors $\mathbf{a}_{1,2}$ (separated by the angle $\theta$) and magnetic unit cells (dashed rectangles).
\label{fig:lattices}}
\end{center}
\end{figure}

We will treat both square and triangular vortex lattices, each with lattice constant $a$, on the same footing. Since each vortex by definition carries magnetic flux equal to a superconducting flux quantum $\Phi_0 = hc/2e$, we can relate the lattice constant $a$ to the magnetic length $\ell_B$ via the angle $\theta$ between primitive translation vectors (see Fig.~\ref{fig:lattices}) as follows:
\begin{align}
a^2 \sin\theta = \frac{hc}{2eB} = \pi \ell_B^2~~,~~\theta = 
    \begin{cases}
        \pi/2 & \mathrm{(square)} \\ 
        \pi/3 & \mathrm{(triangular)}
    \end{cases}. \label{eq:theta}
\end{align}
For the coefficients $C_j$ in Eq.~\eqref{eq:DLLL} we have:
\begin{align}
C_j = C_0 \, e^{-i \pi j^2 \cos\theta} = 
    \begin{cases}
        C_0 & \mathrm{(square)} \\ 
        C_0 \, e^{-i\frac{\pi}{2}j^2} & \mathrm{(triangular)}
    \end{cases}, \label{eq:Cj}
\end{align}
where $C_0 \in \mathbb{R}$ is an energy scale parameterizing the strength of the proximity-induced superconducting pairing. In Fig.~\ref{fig:lattices}, we show the resulting $|\Delta(\rvec)|$ for each respective vortex lattice solution, as well as the corresponding primitive translation vectors and magnetic unit cells \footnote{Equation~\eqref{eq:Cj} can be derived by requiring that $|\Delta(\mathbf{r})|$ be invariant under the magnetic translations $T_1$ and $T_2$ discussed below.}. While this solution for $\Delta(\rvec)$ is, within GL theory, valid exactly at $H_{c2}$, we do expect our qualitative conclusions to hold more generally for fields $H_{c1} \ll B \lesssim H_{c2}$ \footnote{For an extensive review of this ``Ginzburg-Landau'' limit of the Abrikosov vortex lattice, including how to improve upon Eq.~\eqref{eq:DLLL} as one moves away from $H_{c2}$, see Ref.~\cite{rosenstein_ginzburg-landau_2010}.} since our analysis is largely symmetry-based.

We remark that most treatments of Abrikosov vortex lattices in related contexts (see, for example, Refs.~\cite{liu_electronic_2015,murray_majorana_2015,chen_hierarchy_2016}) typically work in the ``London'' limit appropriate for intermediate flux densities, $H_{c1} \lesssim B \ll H_{c2}$. In this regime, the superconducting coherence length $\xi$ is much less than the intervortex separation, i.e., $\xi \ll a$, so that $|\Delta(\rvec)| = \text{const.}$ except near the vortex cores of radius $\sim \xi$; furthermore, the magnetic field is strongly inhomogeneous and is governed by a London-type equation \cite{michael_tinkham_introduction_1996,j_b_ketterson_superconductivity_1999}. Our choice of working in the high-field limit $B \lesssim H_{c2}$ with $a \sim \xi$ is advantageous from a technical standpoint in that it affords us the luxury of working with an \emph{explicit} form for $\Delta(\rvec)$ in the presence of a \emph{constant} magnetic field. We later account for spatial inhomogeneities in the field due to screening, $B(\rvec) = B_0 + \delta B(\rvec)$ with  $\delta B \sim \mathcal{O}(\Delta^2)$, at a phenomenological level by introducing Landau-level broadening via a dispersion relation. By then casting the pairing in a Landau-gauge-based magnetic Bloch basis adapted to $\Delta(\rvec$) (see Sec.~\ref{sec:MBB} below) and subsequently performing Landau-level projection, we circumvent the need for ingenious gauge transformations \cite{liu_electronic_2015,murray_majorana_2015,ariad_how_2018} (see also Ref.~\cite{zocher_topological_2016}).

\subsection{Landau levels of a Rashba spin-orbit coupled 2DEG} \label{sec:RashbaLL}

We now review the Landau-level structure of the single-particle eigenstates which diagonalize $H_\mathrm{2DEG}$ [see Eq.~\eqref{eq:H2DEG}] taking a constant perpendicular magnetic field and zero external potential, $V(\rvec) = 0$ \cite{rashba_properties_1960,bychkov_oscillatory_1984,schliemann_variational_2003,shen_resonant_2004,shen_resonant_2004}. Working in the same Landau gauge ($\mathbf{A}_0 = B x \hat{y}$) which was used to obtain $\Delta(\rvec)$ and defining the standard Landau-level ladder operators $a = \frac1{\sqrt{2}} [ (x / \ell_B - k \ell_B)  + i p_x \ell_B/\hbar]$, with $[a, a^\dagger] = \mathbbm{1}$ and $k$ the plane-wave the momentum in the $y$ direction, the single-particle Hamiltonian for each $k$ reads
\begin{align}
\mathcal{H}_\mathrm{2DEG} = \hbar \omega_c \left[\left(a^\dagger a + \frac12\right) + g_Z \sigma^z + g_R\left( a \sigma^- + a^\dagger \sigma^+ \right) \right] \label{eq:H2DEG_sp}
\end{align}
where $\sigma^{\pm} = (\sigma^x \pm i\sigma^y)/2$ and
\begin{align}
g_R \equiv \frac{\sqrt{2} \alpha_R}{\ell_B \omega_c} = \frac{2 m^2 c}{\hbar e}\left(\frac{\alpha_R}{\sqrt{B}}\right),~~g_Z \equiv \frac{E_Z}{\hbar \omega_c} = \frac{g}{4} \frac{m}{m_e} \label{eq:gRgZ}
\end{align}
parameterize the Rashba and Zeeman coupling strengths, respectively, in units of the cyclotron gap (in the latter case, we have used $E_Z = \frac12 g \mu_B B$ with $g$ the Land\'e $g$-factor and $\mu_B = e \hbar / 2 m_e c$ the Bohr magneton).

\begin{figure}[t]
\begin{center}
\includegraphics[width=0.9\columnwidth]{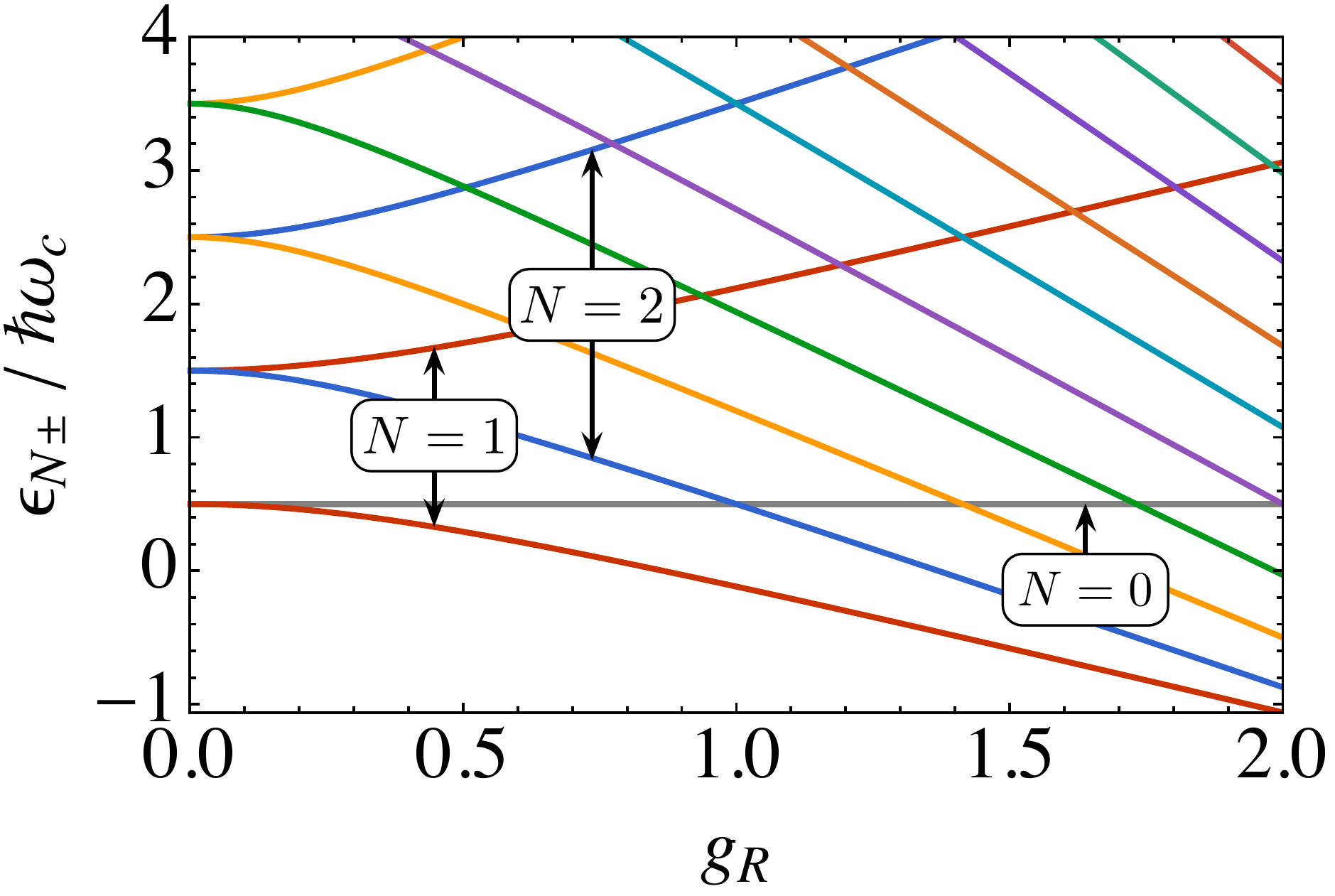}
\caption{Landau-level energies $\epsilon_{N \pm}$ [Eq.~\eqref{eq:epsNpm}] as a function of reduced Rashba coupling $g_R$ [Eq.~\eqref{eq:gRgZ}] evaluated at zero Zeeman coupling, $g_Z = 0$. Levels labeled by $N \geq 1$ are obtained by diagonalizing $\mathcal{H}_N$ [Eq.~\eqref{eq:HN}] in the subspace $\{\ket{N, \uparrow}, \ket{N-1, \downarrow}\}$; the $N = 0$ level corresponds to the unpaired state $\ket{0, \uparrow}$ with energy $\epsilon_0 = \hbar\omega_c(\frac12 + g_Z)$.
\label{fig:RashbaLLspec}}
\end{center}
\end{figure}

The full single-particle Hilbert space is spanned by states $\ket{N, \sigma}$ with (bare) Landau-level index $N = 0, 1, \dots, \infty$ and spin projection $\sigma = \, \uparrow, \downarrow$. It is clear that $\mathcal{H}_\mathrm{2DEG}$ only couples states within spaces $\{\ket{N, \uparrow}, \ket{N-1, \downarrow}\}$. In each such space, given that $N \geq 1$, we are left to contend with
\begin{align}
\mathcal{H}_N = \hbar \omega_c\left[N + g_R \sqrt{N} \sigma^x + \left(1/2 + g_Z\right)\sigma^z \right], \label{eq:HN}
\end{align}
resulting in an energy spectrum given by
\begin{align}
\epsilon_{N \pm} = \hbar \omega_c \left[ N \pm \sqrt{ \left(1/2 + g_Z\right)^2 + N g_R^2} \, \right]. \label{eq:epsNpm}
\end{align}
In addition, there is an unpaired level $\ket{0, \uparrow}$ whose energy $\epsilon_0 = \hbar\omega_c(\frac12 + g_Z)$ is independent of the Rashba coupling. In Fig.~\ref{eq:epsNpm}, we show the evolution of the spectrum as a function of $g_R$ for $g_Z = 0$. Note that $g_R \to 0$ corresponds either to the limit $\alpha_R \to 0$ or $B \to \infty$. Likewise, $g_R \to \infty$ corresponds to either $\alpha_R \to \infty$ or $B \to 0$; in this limit, $\epsilon_{N\pm} \approx \pm \hbar \omega_c \, g_R \sqrt{N}$, which is the Dirac spectrum.

\subsection{Representing superconducting pairing in the Landau and magnetic Bloch bases} \label{sec:MBB}

With the eventual goal of diagonalizing $H_\Delta$ [Eq.~\eqref{eq:HDelta}] upon projection into a given Landau level of $\mathcal{H}_\mathrm{2DEG}$ [Eq.~\eqref{eq:H2DEG_sp}], we aim to express the former in magnetic Bloch bases adapted to the respective symmetries of the square and triangular vortex lattice pairing potentials $\Delta(\rvec)$. Our procedure is completely general in that it applies to arbitrary Landau levels. As a first step, we transform the pairing operator into the Landau basis without Rashba coupling. In Landau gauge, the single-particle wave functions on the infinite plane are $\phi_{mk}(\rvec) = \langle \rvec | m, k \rangle = \frac1{\sqrt{2\pi}}\,\phi_m(x - k \ell_B^2 ) e^{i k y}$ with $\phi_m(x) \equiv \frac{1}{\sqrt{ \ell_B \sqrt{\pi} 2^m m!} }H_m(x / \ell_B) e^{ - \frac{1}{2}(x / \ell_B)^2}$ ($H_m$ being Hermite polynomials). Writing $H_\Delta = \hat{\Delta} + \mathrm{H.c.}$, we have $\hat{\Delta} \equiv \int d^2 r\,\Psi^\dagger_\uparrow(r) \Delta(r) \Psi^\dagger_\downarrow(r) = \sum_{m,n} \int_{k,k'} \Psi_{m\uparrow}^\dagger(k)\,\Delta_{mn}(k, k')\,\Psi_{n\downarrow}^\dagger(k')$, where $\int_k \equiv \int \frac{d k}{2 \pi}$ and $\Psi_{m\sigma}^\dagger(k)$ are the creation operators for the $\phi_{mk}(\rvec)$ orbitals.

The matrix elements in question are thus
\begin{align}
\Delta_{mn}(k, k') \equiv \int d^2 r\,\phi^\ast_{m k}(\rvec) \Delta(\rvec) \phi^\ast_{n k'}(\rvec). \label{eq:DLG_gen}
\end{align}
Using the form of the pairing potential given in Eq.~\eqref{eq:DLLL}, we evaluate this integral in Appendix~\ref{app:LGDelta} and find
\begin{align}
\Delta_{mn}(k, k') = & \sum_j C_j \, \delta(k + k' - k_j) \nonumber \\ & ~~ \times A_{mn} \frac1{\sqrt{2}} e^{-\frac{1}{4} q^2 \ell_B^2 }  H_{m+n}\left(\frac{q \ell_B}{\sqrt{2}}\right), \label{eq:DLG}
\end{align}
where $q \equiv k - k'$ and
\begin{align}
A_{mn} \equiv \frac{(-1)^m}{2^{m+n} \sqrt{m! n!}}.
\end{align}
Of particular importance below are the matrix elements corresponding to $m = n = 0$:
\begin{align}
\Delta_{00}(k, k') = \sum_j C_j \, \delta(k + k' - k_j) \frac1{\sqrt{2}} e^{-\frac{1}{4} (k - k')^2 \ell_B^2 }. \label{eq:D00_Landau}
\end{align}

We next transform from the Landau basis to the magnetic Bloch basis.
Let $T_{1,2}$ be the magnetic translation operators which translate by one primitive translation vector along directions $\avec_{1,2}$ (see Fig.~\ref{fig:lattices}). With one superconducting flux quantum penetrating each unit cell of the vortex lattice (i.e., $\frac{\phi}{\phi_0} = \frac{p}{q} = \frac12$ with $\phi_0 = hc/e = 2\Phi_0$, so that each \emph{magnetic} unit cell contains two vortices, cf.~Fig.~\ref{fig:lattices}), $T_1$ and $T_2$ do not commute with each other but rather satisfy $T_1 T_2 = -T_2 T_1$. We can thus construct a basis---the magnetic Bloch basis---which simultaneously diagonalizes $T_1$, $T_2^2$, as well as the Hamiltonian. This basis is constructed via a linear superposition of Landau orbitals $\ket{N, k_y}$ as follows (see Appendix~\ref{app:MBB} for details):
\begin{align}
& \ket{N, \mathbf{k}} = \sqrt{\frac{Q \ell_B^2}{2 \pi}} \sum_{r=-\infty}^\infty e^{i k_x k_y \ell_B^2} \left(e^{i \kvec \cdot \avec_2}  T_2 \right)^{2 r}  \ket{N, k_y} \label{eq:MagBlochBasis}  \\
 &= \sqrt{\frac{Q \ell_B^2}{2 \pi}} \sum_{r=-\infty}^\infty e^{i k_x ( k_y + r Q ) \ell_B^2 - i 2 \pi r^2 \cos\theta} \ket{N, k_y + r Q}, \nonumber
\end{align}
where we have defined the wave vector $Q \equiv \frac{2 \pi}{a}$ and chosen a normalization $\langle m, \mathbf{k} | n, \mathbf{k}' \rangle = \delta_{mn} \delta^{(2)}(\mathbf{k} - \mathbf{k}')$; the choice of phase factor is motivated by the Landau gauge center of mass relation $\langle x \rangle_{k_y} = k_y \ell_B^2$. Equation~\eqref{eq:MagBlochBasis} encapsulates both the square and triangular vortex lattices through the parameter $\theta = \pi/2, \pi/3$, respectively [cf.~Eq.~\eqref{eq:theta}]; note also that the length scale $Q \ell_B^2 = 2 a \sin\theta$ is lattice-type-dependent.

In this Bloch basis, with associated creation operators $\Psi_{m\sigma}^\dagger(\mathbf{k})$, the lattice momentum is conserved and the pairing operator takes the form
\begin{align}
\hat{\Delta} = \sum_{m,n} \int_\kvec \Psi^\dagger_{m\uparrow}(\mathbf{k}) \Delta_{mn}(\mathbf{k}) \Psi^\dagger_{n\downarrow}(-\mathbf{k}),
\label{eq:Delta_MBB}
\end{align}
where the integral is over the magnetic Brillouin zone (mBZ; see Fig.~\ref{fig:D00}): $\int_\kvec \equiv \int_\mathrm{mBZ} \frac{d^2 k}{(2 \pi)^2}$. It suffices to first focus on the pure LLL matrix elements $\Delta_{00}(\mathbf{k})$. Applying the transformation in Eq.~\eqref{eq:MagBlochBasis} to the Landau basis matrix elements of Eq.~\eqref{eq:D00_Landau} yields
\begin{align}
\Delta_{00}(\mathbf{k}) = \Delta_0 \sum_r e^{-i k_x (2k_y + r Q)\ell_B^2 - \frac{1}{4}(2k_y + r Q)^2 \ell_B^2 + i\pi r^2 \cos\theta}. \label{eq:D00_Bloch}
\end{align}
Here, we have defined $\Delta_0 \equiv C_0 / \sqrt{2}$ and also assumed that $\kvec \in \mathrm{mBZ}$. For evaluation, this expression can be cast in terms of the Jacobi theta function,
\begin{align}
\vartheta_3(z; \tau) \equiv \sum_{n=-\infty}^\infty e^{i\pi \tau n^2 + 2 \pi n i z},
\end{align}
as follows:
\begin{align}
\Delta_{00}(\mathbf{k}) &= \Delta_0 e^{(k_x - ik_y)^2 \ell_B^2 - k_x^2 \ell_B^2} \nonumber \\ & \times \vartheta_3\left[z = (k_x - i k_y)\ell_B\sqrt{\frac{\sin\theta}{\pi}}\,;\,\tau = e^{i \theta}\right]. \label{eq:D00_1}
\end{align}
In Appendix~\ref{app:DeltaDerivatives}, we show how one can obtain arbitrary $\Delta_{mn}$ by taking derivatives of $\Delta_{00}$. Specifically, we find
\begin{align}
\Delta_{mn}(\mathbf{k}) = A_{mn} \, H_{m+n}\left(\frac{i \partial_{k_x}} {\ell_B \sqrt{2}} \right) \Delta_{00}(\mathbf{k}). \label{eq:Dmn}
\end{align}
Note that $\Delta_{00}$ is parity even so that $\Delta_{mn}$ has parity $(-1)^{m+n}$:
\begin{equation}
\Delta_{mn}(-\mathbf{k}) = (-1)^{m+n} \Delta_{mn}(\mathbf{k}).
\end{equation}
These expressions for 
$\Delta_{mn}(\mathbf{k})$ will be used below in Sec.~\ref{sec:PD} to compute the phase diagram in the vicinity of a plateau transition of the superconductor-2DEG hybrid system.

\section{Phase diagram in the vicinity of a quantum Hall plateau transition} \label{sec:PD}

We now envision sweeping the chemical potential $\mu$ \footnote{Experimentally, it will likely be easier to instead tune the strength of the magnetic field $B$ at fixed $\mu$, but for our theoretical analysis here it is conceptually simpler to consider tuning $\mu$ at fixed $B$.} through one of the low-lying Landau levels of $H_\mathrm{2DEG}$, the energies of which are depicted in Fig.~\ref{fig:RashbaLLspec} (for $g_Z = 0$ and no Landau-level broadening). In the limit of large Landau-level separation relative to the pairing gap, 
we can formally project the problem into a single Rashba-coupled Landau level using the machinery developed in Sec.~\ref{sec:setup}. Due to the presence of the vortex lattice, the functions $\Delta_{mn}(\kvec)$ computed above exhibit intricate nodal structures which govern the nature of the plateau transition of interest.

\subsection{Lowest spinful Landau level at $g_R = 0$} \label{sec:spinful}

As a warmup, we first address the nature of the plateau transition upon sweeping the chemical potential through the lowest spinful Landau level in the limit of vanishing Rasbha coupling: $g_R = 0$. Assuming $\hbar\omega_c \gg \Delta_0$, we can project the Hamiltonian into this Landau level (consisting of states $\ket{0,\uparrow}$, $\ket{0,\downarrow}$) leading to a BdG Hamiltonian $H = \int_\kvec \Psi_\kvec^\dagger H_\mathrm{BdG}(\kvec) \Psi_\kvec$ in the basis $\Psi_\kvec^\dagger=[\Psi_{0\uparrow}^\dagger(\kvec), \Psi_{0\downarrow}(-\kvec)]$ with
\begin{align}
H_\mathrm{BdG}(\kvec) &= \left(\begin{array}{cc} E_Z + \varepsilon_\kvec - \mu & \Delta_{00}(\kvec) \\ \Delta_{00}^\ast(\kvec) & E_Z - \varepsilon_\kvec + \mu \end{array}\right). \label{eq:HBdG_0}
\end{align}
Here we have included a phenomenological broadening of the Landau level $\varepsilon_\mathbf{k}$ to account for any  periodic variation in $B(\rvec)$ and $V(\rvec)$ that we have so far neglected ($\varepsilon_\mathbf{k} = 0$ for a completely flat Landau level; see also Sec.~\ref{sec:RashbaPD} below), and the chemical potential is measured relative to $\epsilon_0(g_Z = 0) = \frac{\hbar\omega_c}{2}$. The resulting energy spectrum reads
\begin{align}
E(\mathbf{k}) = E_Z \pm \sqrt{(\varepsilon_\mathbf{k} - \mu)^2 + |\Delta_{00}(\mathbf{k})|^2}.
\end{align}

\begin{figure}[t]
\begin{center}
\includegraphics[width=\columnwidth]{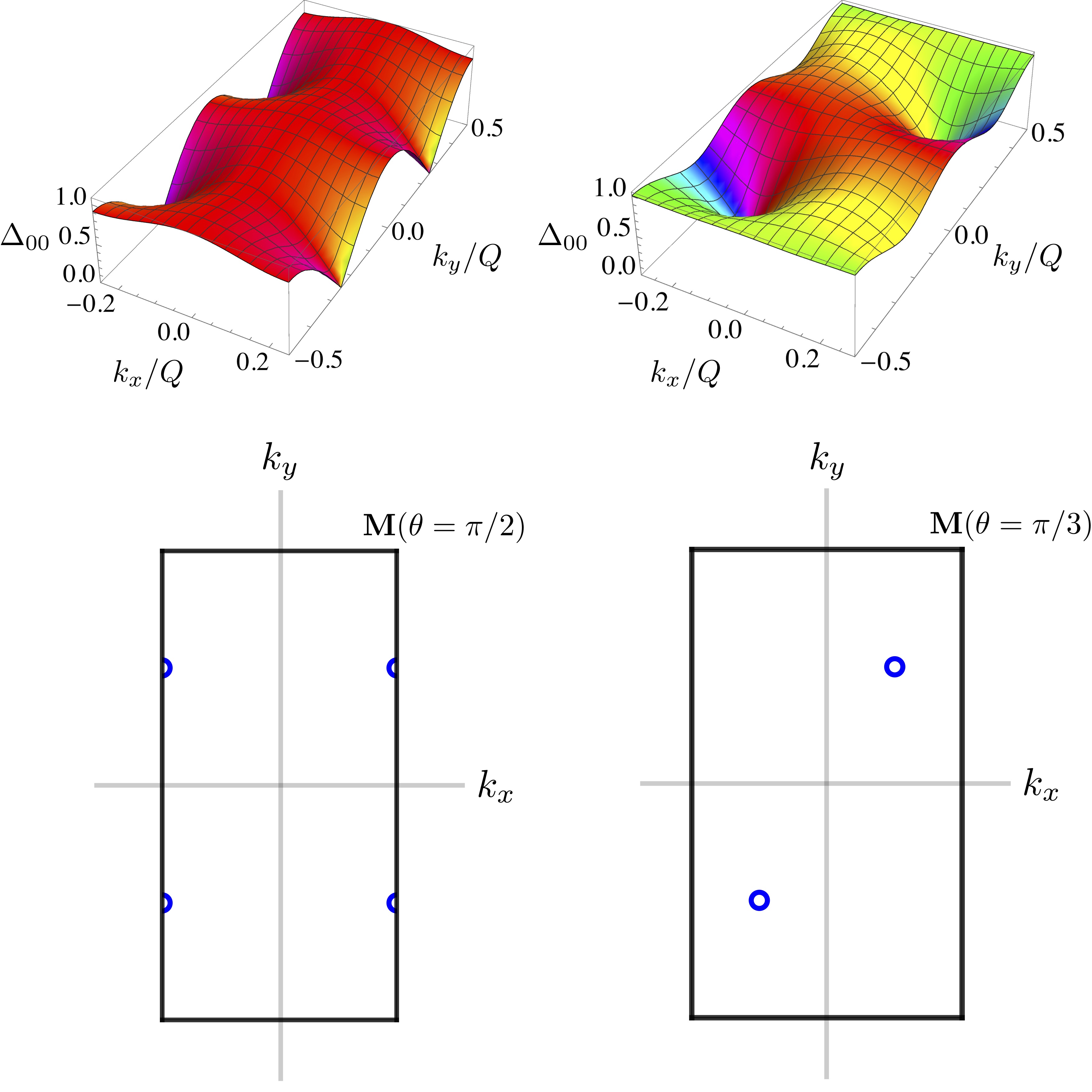}
\caption{Pairing gap function $\Delta_{00}(\kvec)$ for the square (left column) and triangular (right column) vortex lattices. In the top row, the amplitude $|\Delta_{00}(\kvec)|$ is plotted on the $z$ axis, while the phase $\arg\Delta_{00}(\kvec)$ is represented by color. In the bottom panel, we show the locations of the Dirac nodes of $\Delta_{00}(\kvec)$ in the magnetic Brillouin zone (mBZ). For each lattice, the mBZ is rectangular with an $M$ point located at $\mathbf{M}(\theta) = \left( \frac{Q}{4\sin\theta}, \frac{Q}{2} \right)$.
\label{fig:D00}}
\end{center}
\end{figure}

In Fig.~\ref{fig:D00}, we show plots of the gap function $\Delta_{00}(\kvec)$ for the square and triangular vortex lattice pairing potentials as derived in Sec.~\ref{sec:setup}. $\Delta_{00}(\kvec)$ contains Dirac nodes located at momenta $\kvec_\ast$ such that $(k_x - i k_y)\frac{2}{Q}\sin\theta = \left[\left(m + \frac{1}{2}\right) + \left(n + \frac{1}{2}\right)\cos\theta\right] + i\left(n + \frac{1}{2}\right)\sin\theta$ with $m,n \in \mathbbm{Z}$; for each lattice, there are two such nodal points in the mBZ (see Fig.~\ref{fig:D00}).
The phase winding of both nodes in each case have the same, say, ``positive'' chirality.

Further taking the limit $E_Z = 0$ allows a simple understanding of the plateau transition from the viewpoint of a Chern-number-changing transition involving the Dirac nodes of $\Delta_{00}(\kvec)$. We denote the Chern number $C$ as the topological invariant which counts the net number of complex spinless fermions propagating along the edge of the sample \footnote{$C$ is equivalent to the chiral central charge, a quantity which can in principle be directly probed via thermal Hall conductance measurements \cite{banerjee_observation_2018}. The number of Majorana edge modes is $\mathcal{N} = 2C$ (see, for example, Refs.~\cite{qi_topological_2011,qi_chiral_2010}).}. For $E_Z = 0$, the system becomes gapless at (relative) chemical potential $\mu_c = \varepsilon_{\kvec_\ast}$. If the dispersion relation respects symmetries of the vortex lattice (specifically, magnetic translations plus an additional $C_2$ symmetry for the triangular lattice case), then the Fermi surface passes through both nodes simultaneously at which point the spectrum will reveal two Dirac cones. In the \emph{spinful} BdG basis of Eq.~\eqref{eq:HBdG_0}, each of the two Dirac cones will change the Chern number by one; therefore, the total change in Chern number is $\Delta C = 2$. This is the expected behavior: For $\mu < \mu_c$, the system is trivial with $C = 0$, while for $\mu > \mu_c$, we have a topological superconducting analog of two copies of integer quantum Hall with $C = 2$. Note that in the presence of symmetry-breaking perturbations, the single transition will generically be split into an extended $C = 1$ region; however, non-Abelian states with half-integer $C$ cannot appear.

\subsection{Lowest Rashba-coupled Landau level for $g_R > 0$} \label{sec:RashbaPD}

A much more interesting situation arises if we consider the lowest Rashba-coupled ``spinless'' Landau level at finite $g_R$. The pertinent Landau-level states are those with energy
\begin{align}
\epsilon_{1-} = \hbar \omega_c \left[ 1 - \sqrt{ (1/2 + g_Z)^2 +  g_R^2} \right]
\end{align}
obtained by diagonalizing
\begin{align}
\mathcal{H}_1 = \hbar \omega_c\left[\mathbbm{1} + g_R \sigma^x + \left(1/2 + g_Z\right)\sigma^z \right]. \label{eq:H1}
\end{align}
[see Eqs.~\eqref{eq:epsNpm} and \eqref{eq:HN} with $N = 1$.]

We first briefly describe the procedure for projecting the superconducting pairing into a given Rashba-coupled Landau level of $H_\mathrm{2DEG}$. If the solution to the eigenvalue problem of $\mathcal{H}_N$ in the subspace $\{\ket{N, \uparrow}, \ket{N-1, \downarrow}\}$ reads $\mathcal{H}_N \ket{N,\pm} = \epsilon_{N\pm} \ket{N,\pm}$, we can expand the eigenstates in the original Landau basis as
\begin{align}
\ket{N,\pm} = U_{N\uparrow}^{N\pm} \ket{N,\uparrow} + U_{N-1,\downarrow}^{N\pm} \ket{N-1,\downarrow}. \label{eq:dumb}
\end{align}
The associated electron creation operators are thus $\Psi_{N\pm}^\dagger = U_{N\uparrow}^{N\pm} \Psi_{N\uparrow}^\dagger + U_{N-1,\uparrow}^{N\pm} \Psi_{N-1,\downarrow}^\dagger$, which upon inverting allows us to express the ``bare'' electron operators $\Psi_{N\sigma}^\dagger$ in terms of these ``dressed'' operators $\Psi_{N\pm}^\dagger$. After inserting the $\Psi_{N\sigma}^\dagger$ into Eq.~\eqref{eq:Delta_MBB}, it is straightforward to project the pairing operator into any one of the (dressed) Rashba-coupled Landau levels. Since the pairing operator contains all Landau levels, this projection is of course appropriate only when the state in question is sufficiently separated in energy from all other states.

\begin{figure}[t]
\begin{center}
\includegraphics[width=\columnwidth]{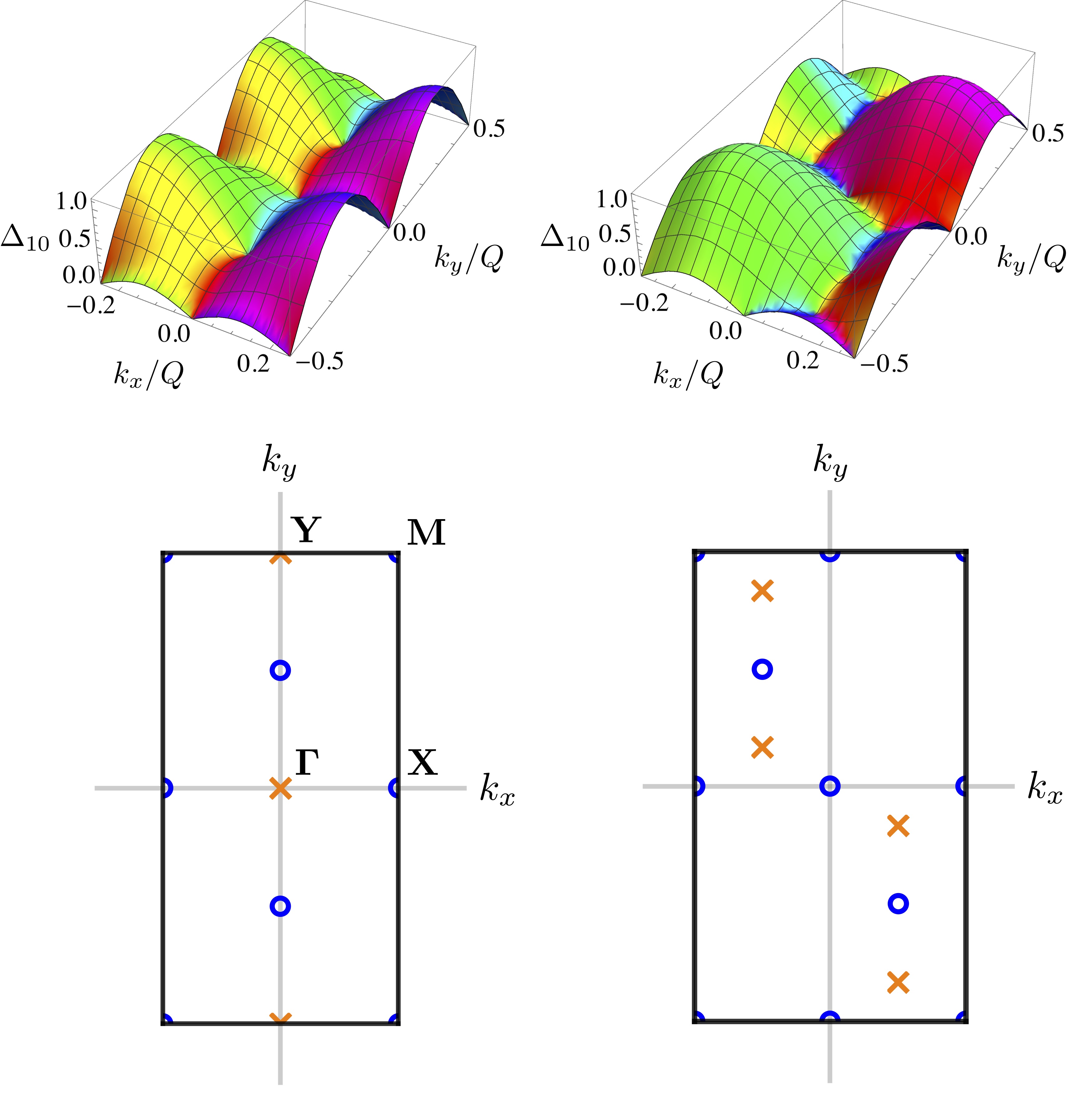}
\caption{Same as in Fig.~\ref{fig:D00}, but now plotting the pairing gap function $\Delta_{10}(\kvec)$ which enters Eq.~\eqref{eq:Hproj} [after being renormalized by $\beta(g_R, g_Z)$, cf.~Fig.~\ref{fig:beta}]. Dirac nodes with positive (negative) phase winding are indicated in the bottom panels by blue circles (orange crosses). As explained in Sec.~\ref{sec:weak}, the gap function for the square lattice case (left panel) gives rise to weak TSC phases.
\label{fig:D10}}
\end{center}
\end{figure}

\begin{figure}[t]
\begin{center}
\includegraphics[width=0.5\columnwidth]{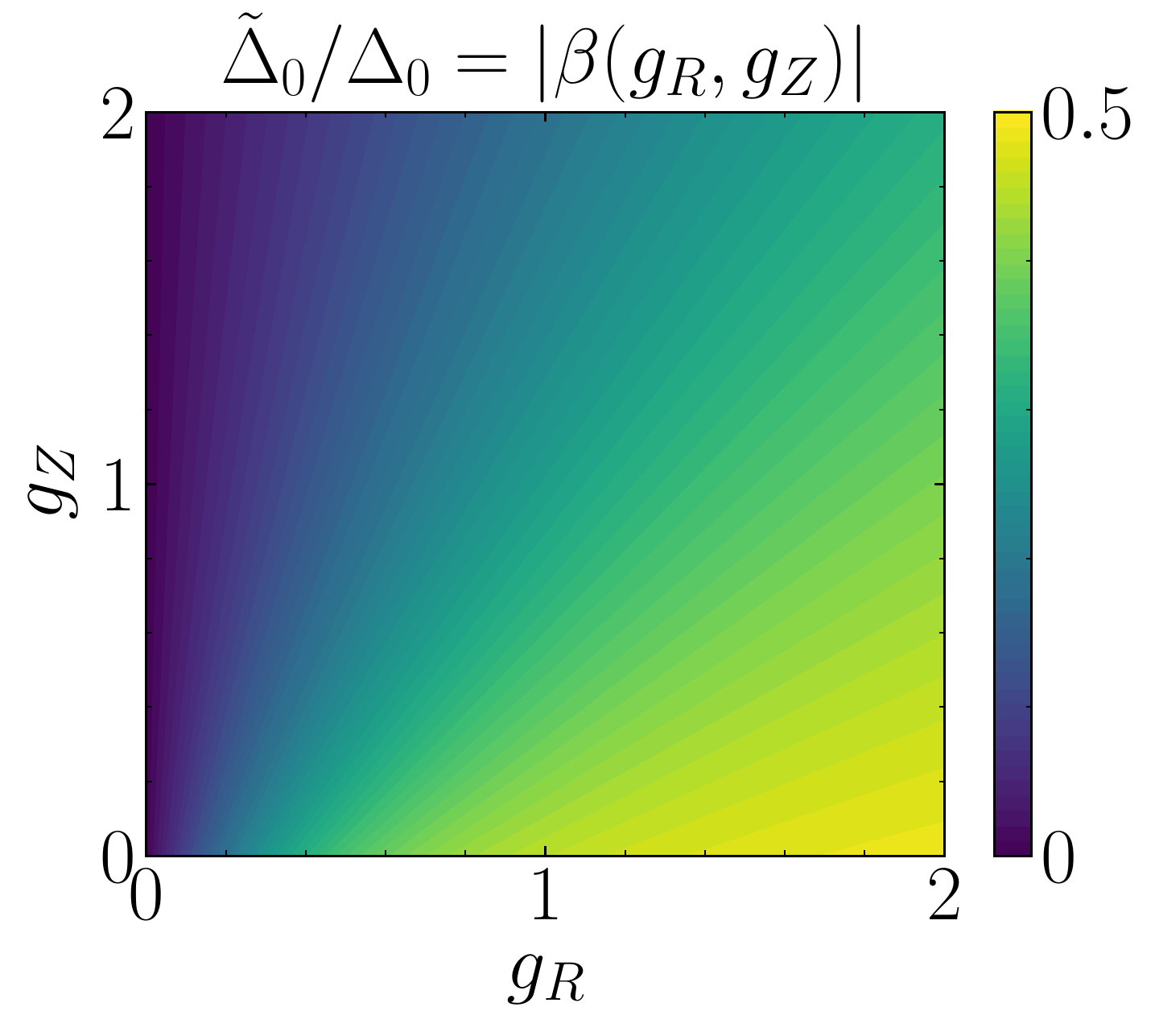}
\caption{Reduced pairing strength $\tilde\Delta_0$ in units of $\Delta_0 \equiv C_0/\sqrt{2}$ [cf.~Eq.~\eqref{eq:Cj}] as a function of $g_R$ and $g_Z$. $\beta(g_R, g_Z)$ is merely the product of wave function amplitudes for the eigenstate of $\mathcal{H}_1$ with energy $\epsilon_{1-}$. Without spin-orbit coupling, the effective pairing strength vanishes, i.e., $\beta(g_R = 0, g_Z) = 0$; while for modest $g_R$, $\tilde\Delta_0 \sim \Delta_0$, e.g., $|\beta(g_R = 1, g_Z = 0)| \approx 0.35$.
\label{fig:beta}}
\end{center}
\end{figure}

If the level closest to the lowest Rashba-coupled Landau level with energy $\epsilon_{1-}$ is the unpaired level with energy $\epsilon_0$ (which occurs for sufficiently small $g_R$ in Fig.~\ref{fig:RashbaLLspec}), projecting entirely into the former can be safely performed provided $\epsilon_0 - \epsilon_{1-} = \hbar \omega_c \left[ \sqrt{ (1/2 + g_Z)^2 +  g_R^2} - \frac{1}{2} \right] \gg \Delta_0$ \footnote{In general, though, we can always consider two nearby levels together if necessary, which would lead to a $4 \times 4$ BdG problem.}. For sufficiently strong Rashba coupling, this condition seems quite reasonable given the high-field limit on which we have based our entire analysis. After introducing a phenomenological broadening via $\epsilon_{1-} \to \epsilon_{1-} + \varepsilon_\kvec$ and measuring $\mu$ relative to $\epsilon_{1-}$, the projected BdG Hamiltonian is $H = \frac12 \int_\kvec \Psi_\kvec^\dagger H_\mathrm{BdG}(\kvec) \Psi_\kvec$, now in the \emph{spinless} BdG basis $\Psi_\kvec^\dagger=[\Psi_{1-}^\dagger(\kvec), \Psi_{1-}(-\kvec)]$, with
\begin{align}
H_\mathrm{BdG}(\kvec) = \left(\begin{array}{cc}\varepsilon_\kvec - \mu & \tilde{\Delta}_{10}(\kvec) \\ \tilde{\Delta}_{10}^\ast(\kvec) & - \varepsilon_\kvec + \mu \end{array}\right).
\label{eq:Hproj}
\end{align}
The corresponding eigenenergies are
\begin{align}
E(\kvec) = \pm \sqrt{(\varepsilon_\kvec - \mu)^2 + |\tilde\Delta_{10}(\kvec)|^2}. \label{eq:Ek}
\end{align}
Here we have defined $\tilde{\Delta}_{10}(\kvec) \equiv \beta(g_R, g_Z) \Delta_{10}(\kvec)$ to be the pairing function $\Delta_{10}(\kvec)$ renormalized by the product of wave function amplitudes $U_{1\uparrow}^{1-} U_{0\downarrow}^{1-} \equiv \beta(g_R, g_Z) \in [-\frac12, 0]$, cf.~Eq.~\eqref{eq:dumb}. In what follows, we quote energies in units of the reduced pairing strength $\tilde\Delta_0 \equiv |\beta(g_R, g_Z)| \Delta_0$ (see Fig.~\ref{fig:beta}). Note that in the limit of vanishing Rashba coupling, the Landau level is completely spin polarized giving $\beta(g_R \to 0, g_Z) \to 0$; thus, as it very well should, the projected $s$-wave pairing function also vanishes in this limit: $\tilde{\Delta}_{10}(\kvec) \to 0$.

The BdG Hamiltonian in Eq.~\eqref{eq:Hproj} looks just like that of a spinless $p + ip$ superconductor \cite{read_paired_2000}, but with a modified pairing potential, $\tilde\Delta_{10}(\kvec)$, and dispersion, $\varepsilon_\kvec$. The former however takes on a nontrivial structure with multiple Dirac nodes of both positive and negative chirality---see Fig.~\ref{fig:D10}. In the bottom panels of Fig.~\ref{fig:D10}, we highlight the precise locations/chiralities of these Dirac nodes: For the square lattice, there are four (two) nodes with positive (negative) chirality, while for the triangular lattice, there are six (four) nodes with positive (negative) chirality \footnote{The latter is consistent with that obtained in Ref.~\cite{zocher_topological_2016}.}.

Now, in this spinless BdG basis, when the Fermi surface passes through a given Dirac node, the Chern number will change by $\Delta C = \pm \frac12$, depending on the chirality of the node. (For reference, a single $p + ip$ superconductor in the continuum has a pairing function $\Delta(\kvec) \sim k_x + i k_y$ with a single Dirac node located at zero momentum, and so for a quadratically dispersing band, the topological transition occurs at $\mu = 0$ \cite{read_paired_2000}: The trivial phase for $\mu < 0$ has Chern number $C = 0$, while the non-Abelian topological phase for $\mu > 0$ has Chern number $|C| = \frac12$ \footnote{The sign of $C$ depends on whether the pairing is $p + ip$ versus $p - ip$.}, indicating the presence of a single chiral Majorana edge mode and concomitant non-Abelian Ising anyonic excitations.)

In the limit of a completely flat Landau level, $\varepsilon_\mathbf{k} = 0$, the chemical potential passes through all nodes of $\tilde\Delta_{10}(\kvec)$ simultaneously. Therefore, the system undergoes a Chern-number-changing transition at $\mu = 0$ (recall here $\mu$ is measured relative to the energy $\epsilon_{1-}$ of the Landau level) with $\Delta C = 1$ for both lattice types \footnote{For the square lattice, $\Delta C = \frac12 (4 - 2) = 1$, while for the triangular lattice $\Delta C = \frac12 (6 - 4) = 1$, where the quantities in parentheses are the total windings ($\#$ positive nodes $-$ $\#$ negative nodes) of the gap function $\Delta_{10}(\kvec)$.}. For $\mu < 0$, the ground state has $C = 0$, while for $\mu > 0$, the system is a topological superconductor (TSC), albeit with $C = 1$ (this phase is adiabatically connected to the $\nu = 1$ integer quantum Hall state).

The $C=0$ and $C=1$ phases can natually be interpreted as the two possible states arising from a Majorana lattice hopping model embedded in a host non-Abelian $p+ip$ superconductor \cite{grosfeld_electronic_2006,kraus_majorana_2011,laumann_disorder-induced_2012,zhou_hierarchical_2013,liu_electronic_2015,murray_majorana_2015}. The latter without a vortex lattice has $C=\frac{1}{2}$, as would arise if we had kept the Zeeman energy but neglected the orbital effect of the magnetic field \cite{sato_non-abelian_2009,sau_generic_2010,alicea_majorana_2010,sau_non-abelian_2010}. Adding orbital effects, the resulting vortex lattice would trap a lattice of Majorana modes, and the total Chern number would include the contribution $C_\chi$ from the band structure of the Majoranas: $C = \frac{1}{2} + C_\chi$. $\mu$ then tunes a Chern-number-changing transition of the Majorana lattice: $C_\chi = -\frac{1}{2} \to \frac{1}{2}$. (As will be discussed, the intermediate non-Abelian phase, $C_\chi = 0$, is forbidden by the magnetic algebra.) However, it appears to us there is an obstruction to determining a microscopic relation between our Landau-level-projected Bloch operators $\Psi_\kvec^\dagger=[\Psi_{1-}^\dagger(\kvec), \Psi_{1-}(-\kvec)]$ and the real-space orbitals of such a Majorana lattice (e.g., the $\gamma_{m, n}$ of Ref.~\cite{affleck_majorana-hubbard_2017}), precisely because the two pictures differ by a non-Wannier-localizable $C = \frac{1}{2}$ phase.

This scenario of a single, direct $C = 0 \to 1$ plateau transition is, however, fine tuned: In general, the Landau level will be broadened by spatially periodic variations in the magnetic field and/or electric potential, so that the Fermi surface will no longer pass through all nodes of $\tilde\Delta_{10}(\kvec)$ at once. We now explore the phase diagram in the vicinity of the above plateau transition in the presence of these more general perturbations. We focus here on the case of the square vortex lattice, as it---in contrast to the triangular lattice---gives rise to \emph{weak} TSC phases which we discuss in Sec.~\ref{sec:weak}. For concreteness, we consider the following dispersion for the Landau level:
\begin{align}
\varepsilon_\kvec &= \varepsilon_\kvec^{(\mathrm{sym})} + \varepsilon_\kvec^{(\mathrm{CDW})}, \label{eq:epsk}
\end{align}
where
\begin{align}
\varepsilon_\kvec^{(\mathrm{sym})} &= -t_b \left(\cos 2 k_x a + \cos 2 k_y a \right), \label{eq:tb} \\
\varepsilon_\kvec^{(\mathrm{CDW})} &= V_\mathrm{CDW} \cos k_y a. \label{eq:cdw}
\end{align}
The first term, $\varepsilon_\kvec^{(\mathrm{sym})}$, represents a phenomenological broadening of the Landau level which preserves all symmetries of the vortex lattice, namely magnetic translations and spatial rotations; the corresponding bandwidth is $t_b$. The second ``charge density wave'' (CDW) term, $\varepsilon_\kvec^{(\mathrm{CDW})}$, arises from a perturbation of period $2a$ in the $x$ direction which doubles the unit cell of the original square lattice and dimerizes the Majorana modes bound to the two corresponding vortices. This unidirectional superlattice perturbation, in contrast, explicitly breaks the magnetic translation symmetry (as well as the $C_4$ rotational symmetry).

\begin{figure}[t]
\begin{center}
\includegraphics[width=0.8\columnwidth]{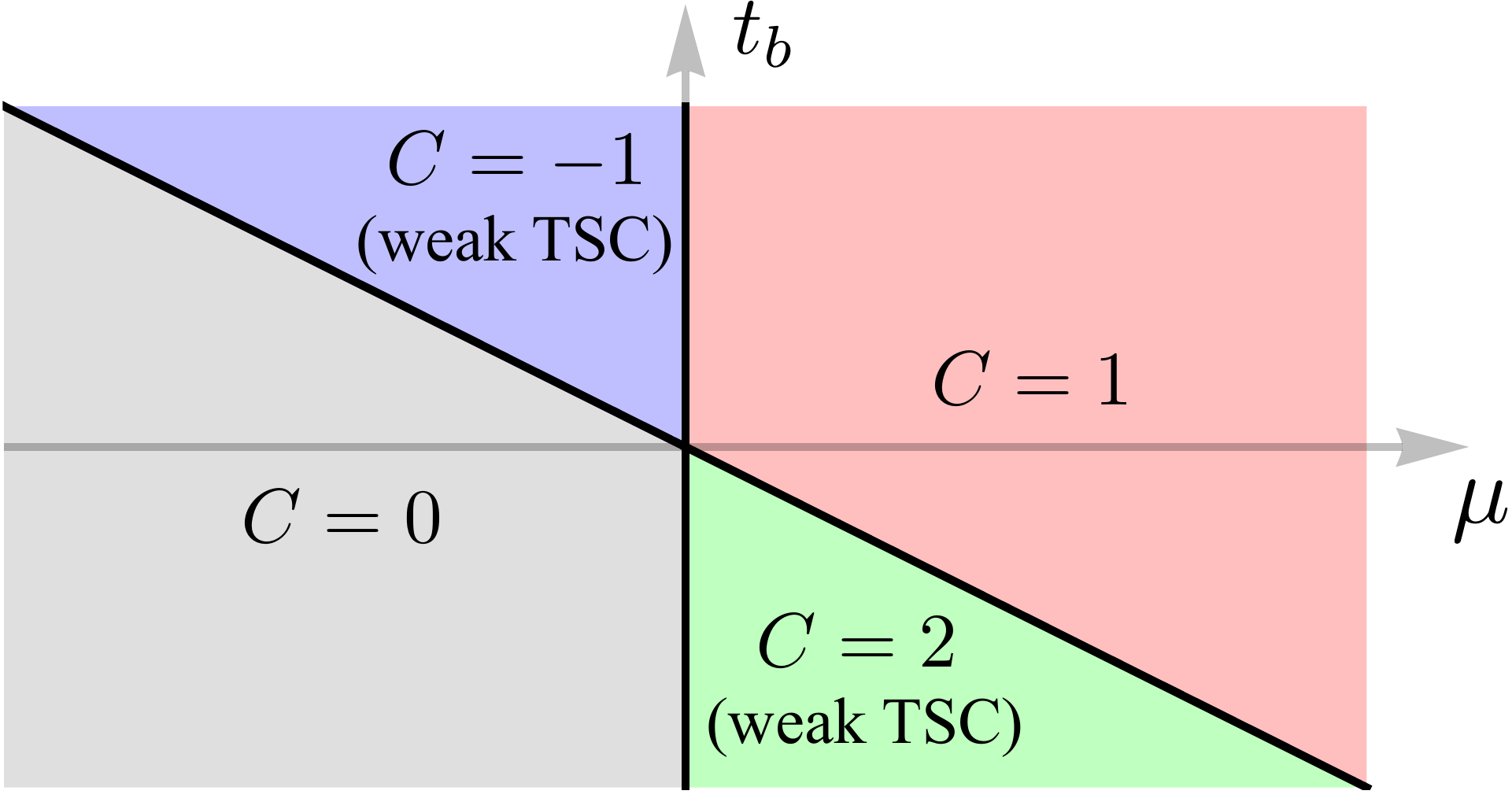}
\caption{Phase diagram in the vicinity of the plateau transition for the square vortex lattice encountered upon sweeping the chemical potential $\mu$ through the lowest Rashba-coupled Landau level of energy $\epsilon_{1-}$ ($\mu$ is measured relative to $\epsilon_{1-}$). Here, $t_b$ is the bandwidth of the Landau level [see Eq.~\eqref{eq:tb}], and the magnetic-symmetry-violating term $V_\mathrm{CDW} = 0$. For a completely flat Landau level ($t_b = 0$), there is a direct Chern number $C = 0 \to 1$ transition at $\mu = 0$, while at finite $t_b$ the transition splits into intermediate $C = -1$ or $2$ states. All phases consist of Majorana lattices; in particular, the $C = -1, 2$ states are weak TSCs (see Sec.~\ref{sec:weak}).
\label{fig:PDtb}}
\end{center}
\end{figure}

As can be gleaned by inspecting the bottom-left panel of Fig.~\ref{fig:D10}, all positive-winding nodes (blue circles) of $\tilde\Delta_{10}(\kvec)$ can be related by some combination of the magnetic translations as well as $C_4$ rotations. (For the former, recall that since $T_1 T_2 = -T_2 T_1$ and $T_1\ket{\kvec} = e^{-i k_y a} \ket{\kvec}$, we can identify $T_2 \ket{\kvec} \sim \ket{\kvec + (0, \frac{Q}{2})}$.) Similarly, the magnetic algebra alone relates the two negative-winding nodes (orange crosses). Therefore---since $T_1$, $T_2$, and $C_4$ all commute with the Hamiltonian---if these symmetries are preserved, the Fermi surface must pass through all four of the positive-winding nodes simultaneously thereby changing the Chern number by $\Delta C = \frac12 \cdot 4 = 2$, and similarly for the two negative-winding nodes with $\Delta C = -\frac12 \cdot 2 = -1$. In general, however, these two transitions need no longer happen at the same chemical potential.

\begin{figure}[t]
\begin{center}
\includegraphics[width=0.65\columnwidth]{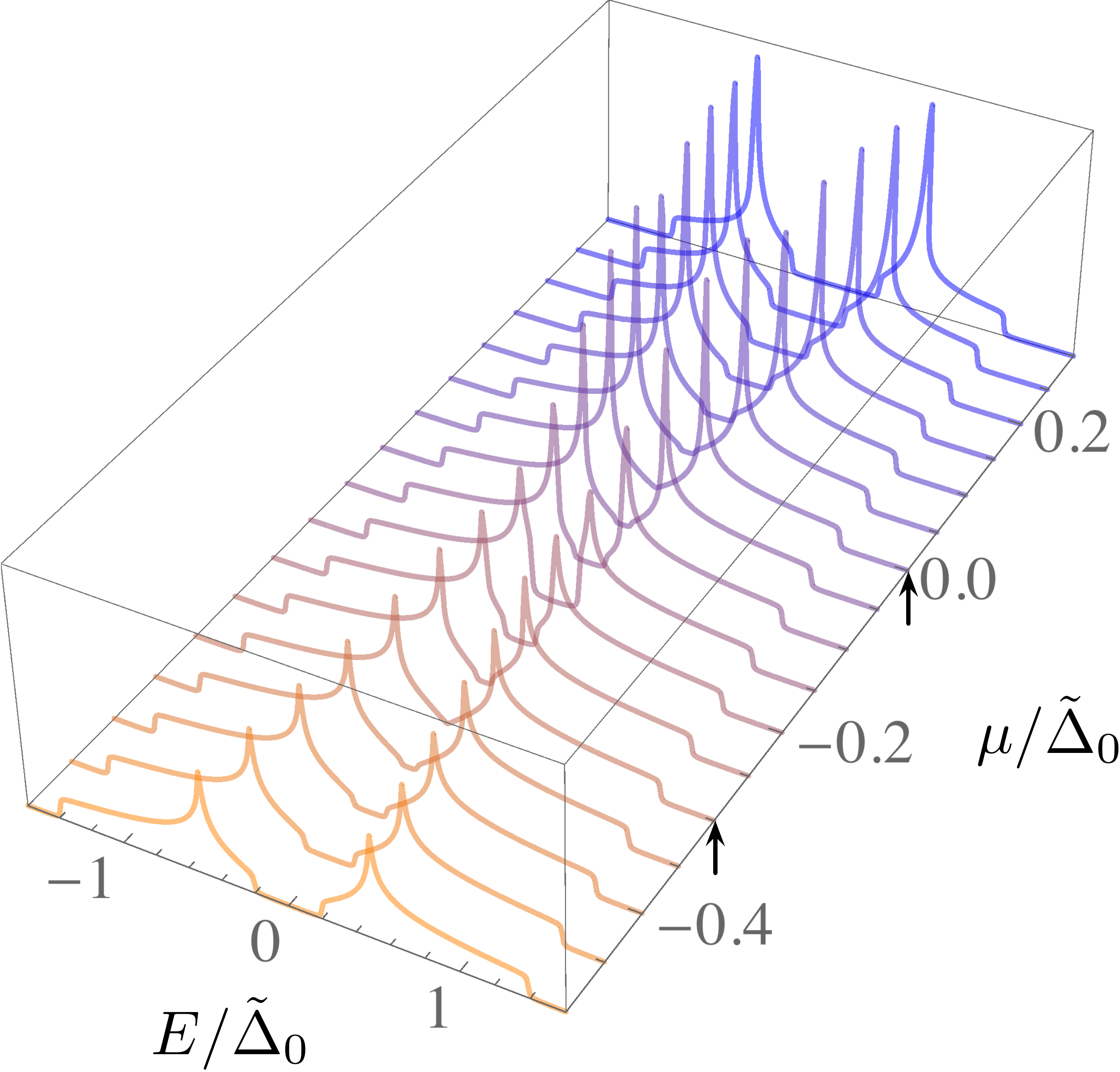}
\caption{Total density of states, $\mathrm{DOS}(E)$, of the Landau-level projected BdG Hamiltonian, Eq.~\eqref{eq:Hproj}, versus chemical potential $\mu$ at fixed Landau-level bandwidth $t_b = 0.15\tilde\Delta_0 > 0$ with $V_\mathrm{CDW} = 0$. The black arrows indicate the critical points $\mu_{c1} = -2t_b$ and $\mu_{c2} = 0$.
\label{fig:DOS}}
\end{center}
\end{figure}

One such example of a symmetry-preserving perturbation is the Landau-level broadening term $\varepsilon_\kvec^{(\mathrm{sym})}$ introduced above. The phase diagram in the space $t_b$ vs $\mu$ (at $V_\mathrm{CDW} = 0$) is shown in Fig.~\ref{fig:PDtb}. For $t_b > 0$, the Chern number sequence upon sweeping $\mu$ through the Landau level goes $C = 0 \to -1 \to 1$, while for $t_b < 0$, it is $C = 0 \to 2 \to 1$ (negative-winding nodes get swept through first in the former case, positive-winding ones first in the latter case). Interestingly, the intermediate integer Chern $C = -1, 2$ phases are actually \emph{weak} TSCs, as we elaborate below in Sec.~\ref{sec:weak}.

Figure~\ref{fig:DOS} shows the total density of states (DOS) per magnetic unit cell,
\begin{align}
\mathrm{DOS}(E) = \int_\kvec \delta[E - E(\mathbf{k})], \label{eq:DOS}
\end{align}
[with $E(\mathbf{k})$ given by Eq.~\eqref{eq:Ek}] versus chemical potential at $t_b = 0.15\tilde\Delta_0 > 0$. The first transition (with $\Delta C = -1$) upon increasing $\mu$ occurs at $\mu_{c1} = -2 t_b = -0.3 \tilde\Delta_0$, while the second transition (with $\Delta C = 2$) occurs at $\mu_{c2} = 0$. At each transition (see arrows in Fig.~\ref{fig:DOS}), there is a clear Dirac signature in the DOS at $E = 0$. Because we have projected to a single Landau level, there is one state per magnetic unit cell (area $2a^2$), or equivalently, $\frac12$ states (i.e. a Majorana) per vortex (area $a^2$): $\int dE\,\mathrm{DOS}(E) = \frac12 \frac1{a^2}$. Unfortunately, such total number of states per unit area is not directly accessible with STM, which measures differential conductance, and thus the density of states only up to a nonuniversal prefactor. Still, as discussed in Sec.~\ref{sec:intro}, the Landau-level limit precludes other low-energy states in the vortex cores, which should be verifiable experimentally.

\begin{figure}[t]
\begin{center}
\includegraphics[width=\columnwidth]{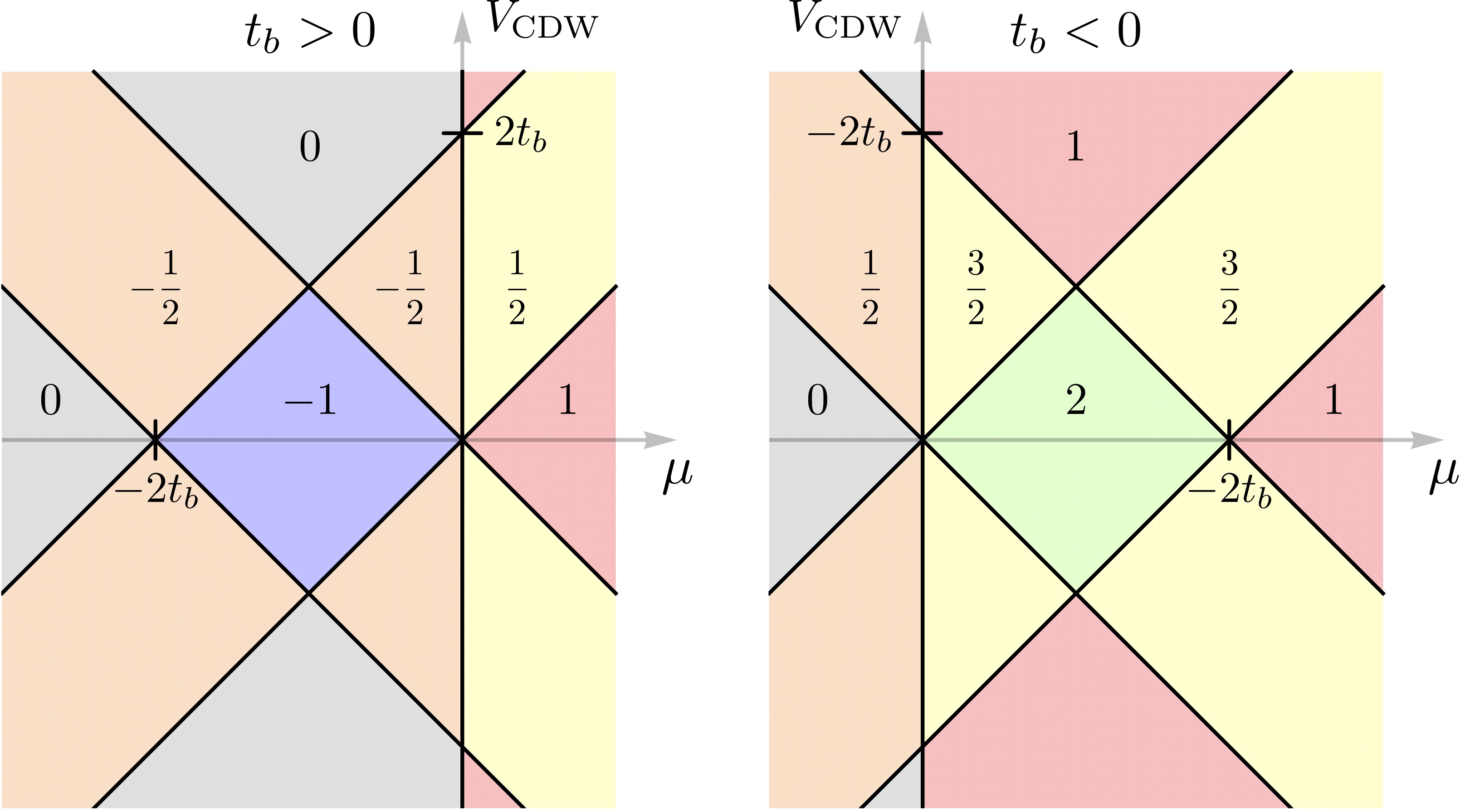}
\caption{Phase diagram near the plateau transition for the square lattice, now including finite $V_\mathrm{CDW}$. Since this term breaks the magnetic translation symmetry [see Eq.~\eqref{eq:cdw}], the transitions in Fig.~\ref{fig:PDtb} get split into multiple transitions involving intermediate non-Abelian states with half-integer $C$, as labeled in the diagram. The left and right panels correspond to different signs of $t_b$, and the phase diagram is symmetric under $V_\mathrm{CDW} \leftrightarrow -V_\mathrm{CDW}$.
\label{fig:PDcdw}}
\end{center}
\end{figure}

We make the following observation: Because the magnetic symmetry relates each node to a partner of the same chirality, its presence prohibits the realization of non-Abelian topological states with half-integer Chern number. In other words, to obtain bona fide non-Abelian topological states, it is thus necessary to introduce perturbations which \emph{violate} the magnetic translation symmetry (see also Ref.~\cite{zocher_topological_2016} as well as Ref.~\cite{jeon_topological_2018}, where in the latter case the considered pairing potential itself breaks the magnetic symmetry). For illustrative purposes and for use in Sec.~\ref{sec:disorder}, we consider as a concrete example the unit-cell-doubling CDW term in Eq.~\eqref{eq:cdw}. The obtained phase diagram as a function of $V_\mathrm{CDW}$ and $\mu$ at fixed $t_b$ is shown in Fig.~\ref{fig:PDcdw} (the two panels correspond to positive and negative $t_b$). Now, indeed the individual phase transitions from Fig.~\ref{fig:PDtb} split into multiple transitions involving half-integer $\Delta C$ and thus intermediate extended non-Abelian states with $C = \pm\frac12, \frac32$.
The extent of these phases, as well as their excitation gaps, are both set by the strength of the perturbation $V_\mathrm{CDW}$. Note also that the phase diagram in Fig.~\ref{fig:PDcdw} is symmetric upon taking $V_\mathrm{CDW} \leftrightarrow -V_\mathrm{CDW}$, although the half-integer states related by flipping the sign of $V_\mathrm{CDW}$ differ by a \emph{weak} topological index described in Sec.~\ref{sec:weak}. We will return to this point in Sec.~\ref{sec:disorder} when discussing the effects of disorder on this Majorana lattice system.

While in the above we have chosen to explicitly work out the case of the square vortex lattice, similar considerations apply equally well to the case of the triangular lattice \cite{zocher_topological_2016}. As is clear from the right panels of Fig.~\ref{fig:D10}, half-integer Chern phases are again precluded provided that magnetic translations, as well as an additional $C_2$ (i.e., $\kvec \leftrightarrow -\kvec$) symmetry, are preserved. One important difference, however, is that now intermediate integer Chern phases are no longer weak TSCs (see Sec.~\ref{sec:weak} below). Finally, we note that the above procedure also carries over to plateau transitions involving higher Landau levels, with the only difference being the gap function input to Eq.~\eqref{eq:Hproj}. For example, for the Landau level at energy $\epsilon_{1+}$, one would work again with $\Delta_{10}(\kvec)$, so in that case the above results apply directly. In general, for levels at $\epsilon_{N\pm}$, one needs to contend with the gap function $\Delta_{N,N-1}(\kvec)$.

\section{Intermediate integer Chern phases as weak topological superconductors} \label{sec:weak}

\begin{figure}[t]
\begin{center}
\includegraphics[width=0.6\columnwidth]{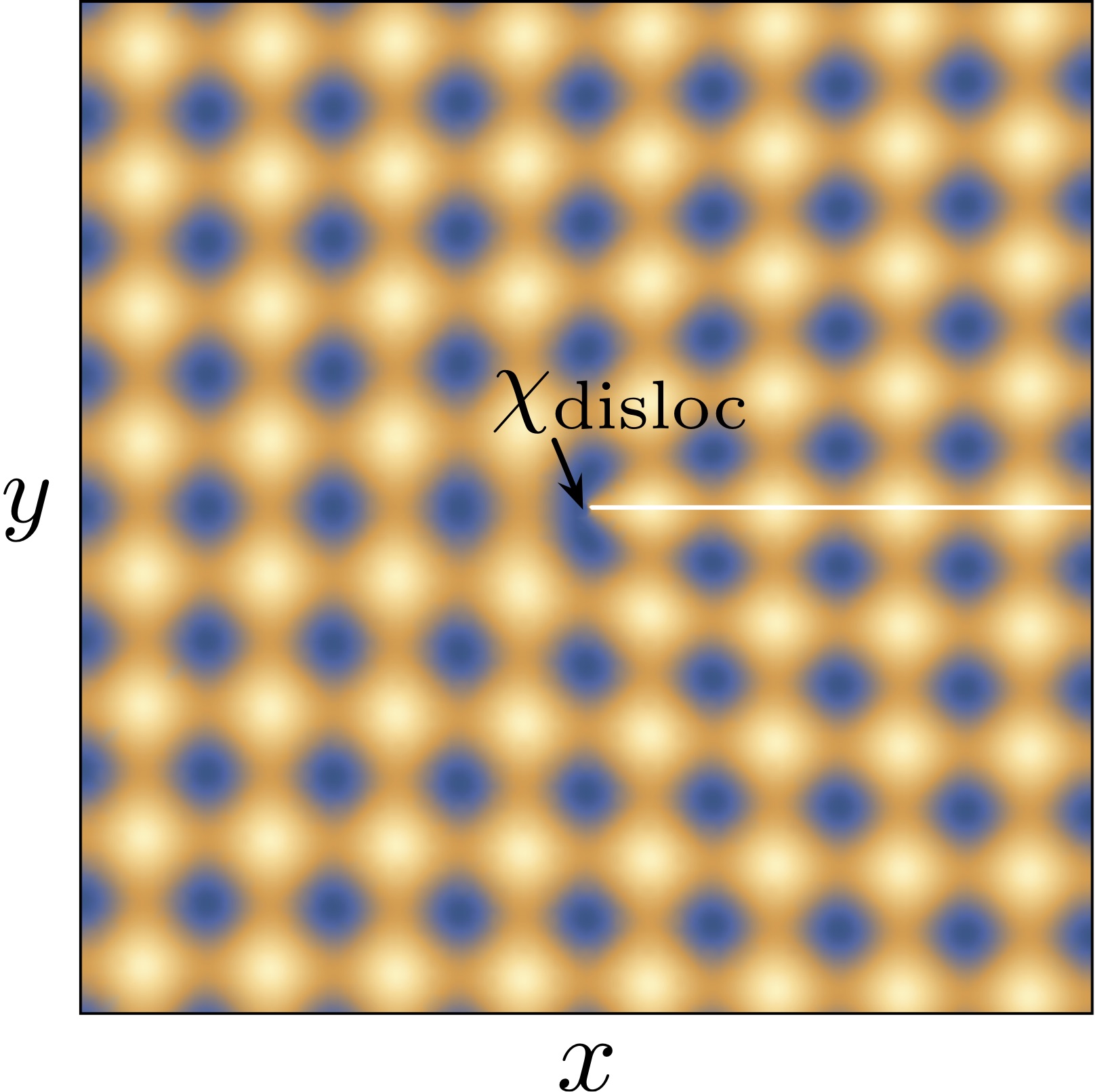}
\caption{Unpaired Majorana zero mode $\chi_\mathrm{disloc}$ occurring at the core of a lattice dislocation in the intermediate $C=-1,2$ phases of the square vortex lattice (see Figs.~\ref{fig:PDtb} and \ref{fig:PDcdw}). The model pairing potential $|\Delta(\rvec)|$ shown here is merely meant to be illustrative and, in contrast to Fig.~\ref{fig:lattices}, is not an accurate rendition of the actual form, Eq.~\eqref{eq:DLLL}, used throughout.
\label{fig:disloc}}
\end{center}
\end{figure}

Weak 2D topological phases arise when an array of 1D topological phases are ``stacked'' side-by-side along either columns or rows. More formally, for a system with a Bravais lattice generated by $T_{1,2}$, we can associate weak indices $\nu_{1,2}$ according to whether 1D topological phase $\nu_i$ is stacked along direction $T_i$ \cite{teo_existence_2013, cheng_translational_2016}. In our context, the relevant 1D phase is the 1D Kitaev chain, which has a Majorana edge state \cite{Kitaev01_PhysU_44_131}. The Kitaev chain is the nontrivial $\nu_i = -1$ element of a $\nu_i \in \mathbb{Z}_2$ classification in the absence of further protecting symmetries. The stacking picture makes evident two physical signatures of a weak TSC. First, because each constituent 1D chain carries a localized edge state, the boundary of the 2D system will host an array of edge states which hybridize into a 1D gapless mode. This gapless mode is stable so long as the translation symmetry of the boundary is preserved. Second, if a dislocation is introduced in the bulk, the core of the dislocation will carry the same zero mode as the edge of the 1D topological phase. In the context of our system, if the realized phase is a weak TSC then a dislocation in the superconducting vortex lattice will carry an unpaired non-Abelian Majorana zero mode, as shown in Fig.~\ref{fig:disloc}. We now show that the intermediate-$\mu$ phases of the square lattice ($C = -1$ or $C = 2$ in Fig.~\ref{fig:PDtb}) is such a weak TSC.

To compute the weak invariant \cite{teo_existence_2013}, consider wrapping the model onto a circumference $L_y$ cylinder. According to the stacking picture, viewed as a 1D system the cylinder is equivalent to $L_y$ copies of a 1D superconductor with index $\nu_1$ each; the resulting 1D invariant is $\nu_{1D} = \nu_1^{L_y}$. So we can take $L_y = 1$ and thereby read off $\nu_1$ by computing the 1D strong invariant. Fixing $L_y = 1$ amounts to projecting onto $k_y = 0$, so the resulting 1D Hamiltonian is
\begin{align}
H_{\textrm{1D}}(k) &= H_{\textrm{BdG}}(k_x = k, k_y = 0)
\end{align}
where $H_{\textrm{BdG}}(\kvec)$ is defined in Eq.~\eqref{eq:Hproj}. The corresponding 1D dispersion is $\varepsilon_k = \varepsilon_{\kvec=(k,0)}$, cf.~Eq.~\eqref{eq:epsk}. To compute $\nu_{1D}$, we then appeal to Kitaev's weak-pairing criteria for an inversion-symmetric TSC~\cite{Kitaev01_PhysU_44_131}. In the absence of pairing, let $\nu(k) = (-1)^{N_{\textrm{occ}}(k)}$ denote the parity of the number of bands below the Fermi surface at momentum $k$; then $\nu_{1D} = \nu(0) \nu(\frac{\pi}{2 a})$ (corresponding to the $\Gamma$ and $X$ points, respectively, in the bottom-left panel of Fig.~\ref{fig:D10}). For $\mu \ll \varepsilon_k$, all states are empty and so $\nu_{1D} = 1$ (trivial); likewise for $\mu \gg \varepsilon_k$, all states are full and $\nu_{1D} = 1$. But for $\textrm{min}\,{\varepsilon_k} < \mu < \textrm{max}\,{\varepsilon_k}$, one of $k = 0, \frac{\pi}{2a}$ will be empty while the other is occupied [which depends on the shape of the dispersion, e.g., the sign of $t_b$ in Eq.~\eqref{eq:tb}]. In this intermediate regime, $\nu_{1D} = -1$ and so the 2D system is a weak TSC with $\nu_1 = \nu_2 = -1$ ($\nu_2$ is identical from $C_4$ symmetry). The transition points are of course precisely where the Fermi surface passes through $k = 0, \frac{\pi}{2a}$ and the superconductor becomes gapless due to the nodes in the pairing $\tilde\Delta_{10}(\kvec)$ at $\Gamma$ and $X$.

The triangular lattice, on the other hand, does not have an intermediate weak TSC, because the weak invariant is incompatible with the $C_3$ point-group symmetry. Nonetheless, the identification of such weak TSC phases for the square lattice, as demonstrated above, further bolsters our argument for the presence of a Majorana lattice in this system.

\section{Effects of disorder} \label{sec:disorder}

Finally, we comment at a qualitative (and rather speculative) level on the effects of disorder, which will naturally arise from imperfections in the vortex lattice. The analogous problem has been addressed previously in Refs.~\cite{kraus_majorana_2011,laumann_disorder-induced_2012} in the context of disordered Majorana lattice hopping models (on the triangular lattice). In the region of the phase diagram with a large gap (e.g., $\mu$ large or small), the disorder will have little impact. However, as $\mu$ approaches the $C$-changing transition, the width of the disorder will become comparable to the gap. We argue that disorder will broaden the transition into a region of ``thermal Majorana metal'' \cite{kraus_majorana_2011,laumann_disorder-induced_2012} so long as the magnetic translation symmetry is preserved \emph{on average}. Consider, for example, the critical point at $\mu_{c1} = -2 t_b$ in the left panel of Fig.~\ref{fig:PDcdw}. Local variations in $\mu$ and $V_{\textrm{CDW}}$ will nucleate domains of the four competing phases. If all are present, there is a gapless intermediate phase because their boundaries carry a chiral Majorana mode, resulting in a gapless network model.

A more delicate possibility, however, is the presence of an intermediate regime in which only regions of $C= -\frac{1}{2}$ appear (for example, if the variance in $V_{\textrm{CDW}}$ is large compared to that in $\mu$). More precisely, there are four such $C = -\frac{1}{2}$ phases corresponding to $\varepsilon_{\kvec=\mathbf{Q}}^{(\mathrm{CDW})} \lessgtr 0$ at wavevectors $\mathbf{Q} = (0, \frac{\pi}{a}), (\frac{\pi}{a}, 0)$, which are related by $T_1, T_2$, and $C_4$. Accordingly, we can label them with a $\mathbb{Z}_4$ index $\psi = e^{i \theta}$ following the order
$
\varepsilon_{\kvec=(0, \pi/a)}^{(\mathrm{CDW})} > 0,  \varepsilon_{\kvec=(\pi/a, 0)}^{(\mathrm{CDW})} > 0, \varepsilon_{\kvec=(0, \pi/a)}^{(\mathrm{CDW})} < 0, \varepsilon_{\kvec=(\pi/a, 0)}^{(\mathrm{CDW})} < 0
$.
While all are strong $C=-\frac{1}{2}$ TSCs, inspection of their band structures verifies that the four phases differ by \emph{weak} indices. Specifically, the two phases related by $T_1$ ($\psi = \pm 1$) differ by $\nu_2$, and the two phases related by $T_2$ ($\psi = \pm i$) differ by $\nu_1$.  Thus, for example, a domain wall between $\psi = \pm 1$ which locally preserves the $T_2$ symmetry will have a low-energy mode.

In the Majorana lattice picture, the situation is directly analogous to the $C_4$-related columnar valence-bond-solid (VBS) dimerization patterns of a spin-$\frac12$ antiferromagnet, with a nearest-neighbor Majorana ``pairing''  (e.g., $i \gamma_{x, y} \gamma_{x, y+1}$) playing the role of a singlet. Just as vortices in a VBS pattern carry an unpaired $S=\frac12$ moment \cite{levin_deconfined_2004}, here we expect vortices in $\psi$ will carry a Majorana mode. Thus, disorder will nucleate a dilute random lattice of Majorana modes which will hybridize into the Majorana version of the random singlet phase. Similar physics was investigated recently in the spin-$\frac12$ case \cite{kimchi_valence_2018}, where a power-law spectrum of low-energy modes was predicted to arise from a disordered version of the Lieb-Schultz-Mattis theorem. It would be interesting to extend their quantitative predictions to the present Majorana case \cite{hsieh_all_2016} since such power laws might be observed in STM spectroscopy, although interacting phases may intervene instead. 

\section{Discussion} \label{sec:discussion}

The quantum Hall effect and superconductivity have traditionally been thought to be largely incompatible: the strong magnetic fields required for the former will generally kill the latter. However, for near-term superconductor-2DEG heterostructure devices, the marriage is not entirely unreasonable if one can choose a 2DEG with a sufficiently small effective mass and a superconductor with a sufficiently large upper critical field---then one can in principle attain large cyclotron gaps without completely destroying proximity-induced superconducting pairing. We have analyzed a physically reasonable limit of this scenario in which an $s$-wave superconductor at fields near its upper critical field $H_{c2}$---and exhibiting the corresponding Abrikosov vortex lattice---is proximity coupled to a strongly spin-orbit coupled 2DEG in the quantum Hall regime. The Landau-level wave functions are \emph{natively} endowed with both spin components by the spin-orbit coupling, thereby allowing $s$-wave superconductivity to give rise to effective \emph{spinless} superconductivity when projected into a single Landau level. This is somewhat analogous to the usual Zeeman-based recipe \cite{sato_non-abelian_2009,sau_generic_2010,alicea_majorana_2010,sau_non-abelian_2010} of engineering spinless $p$-wave superconductivity \cite{read_paired_2000} with the exact same ingredients; however, now orbital effects of the magnetic field dictate the physics, while the Zeeman effect plays essentially no role.

All realized states involve a lattice of Majorana modes. In fact, in this single Landau-level limit of the problem, simple state counting alone tells us that each vortex must necessarily harbor a Majorana mode. However, the symmetries of the underlying vortex lattice preclude the presence of any non-Abelian states with half-integer Chern number $C$ (an additional vortex would thus not bind a Majorana zero mode).
While additional perturbations such as a superlattice modulation could in principle give rise to bona fide non-Abelian states (see Fig.~\ref{fig:PDcdw} and Refs.~\cite{zocher_topological_2016,jeon_topological_2018}), for topological quantum computing purposes, perhaps the most intriguing application of our results involves the \emph{weak topological superconducting} states emerging near the plateau transition for the square vortex lattice at intermediate $\mu$. Can one experimentally pattern a square vortex lattice with intentionally introduced lattice dislocations as a means of engineering the 1D Kitaev model site-by-site \footnote{This is somewhat reminiscent of the strategy proposed in Ref.~\cite{su_andreev_2017} using 1D quantum dot arrays.}? Tuning the system to the weak TSC phase then gives bona fide \emph{Majorana zero modes} at the cores of the dislocations. Even in the absence of such applications though, our proposal for engineering a Majorana lattice deep in the quantum Hall limit seems experimentally appealing.

Finally, adding strong electron-electron interactions to this problem leads to a number of interesting open questions. For example, will the inclusion of interactions \emph{spontaneously} split the integer $\Delta C$ transitions into intermediate non-Abelian states with half-integer $C$? Furthermore, the presence of strong interactions in the 2DEG should be a promising means for engineering \emph{interacting} Majorana lattice models, as have begun to be discussed only recently \cite{rahmani_emergent_2015,rahmani_phase_2015,chiu_strongly_2015,affleck_majorana-hubbard_2017,wamer_renormalization_2018,rahmani_majorana-hubbard_2018,ware_ising_2016}. More generally, it very interesting to contemplate what kinds of ground states could emerge upon adding proximity-induced superconducting pairing to interacting \emph{fractional} quantum Hall states. The near-term experimental setup proposed herein gives new motivation for attacking this challenging problem, both theoretically and numerically.

\acknowledgments

We thank Mallika Randeria and Yonglong Xie for valuable discussions. A.Y.~acknowledges support from the Gordon and Betty Moore Foundation as part of EPiQS initiative (GBMF4530), DOE-BES grant DE-FG02-07ER46419, and NSF-MRSEC programs through the Princeton Center for Complex Materials DMR-142054, NSF-DMR-1608848.

\appendix

\section{Calculational details pertinent to Sec.~\ref{sec:setup}} \label{app:details}

\subsection{Evaluating Landau gauge pairing matrix elements $\Delta_{mn}(k, k')$ [Eq.~\eqref{eq:DLG}]} \label{app:LGDelta}

After performing the integration over $y$ in Eq.~\eqref{eq:DLG_gen}, we need to evaluate an integral of the form (setting $\ell_B = 1$)
\begin{align}
G_{mn}(q) = \frac{1}{\sqrt{\pi 2^{m+n} m! n! }} & \int dx \,H_m(x - q/2) H_n(x + q/2) \nonumber \\ & \times e^{- \frac{1}{2}(x - q/2)^2 -  \frac{1}{2}(x + q/2)^2 - x^2}.
\end{align}
First, note the generating function expansion
\begin{align}
e^{2 t x - t^2} = \sum_n  H_n(x) \frac{t^n}{n!}\,, ~~ \partial_t^n e^{2 t x - t^2}|_{t=0} =  H_n(x).
\end{align}
Thus, it will be sufficient to evaluate
\begin{align}
O(t_1, t_2) &= \int dx \, e^{2 t_1 (x - q/2) - t_1 ^2 - 2 t_2 (x + q/2) - t_2 ^2} \nonumber \\ & ~~~~~ \times e^{-\frac{1}{2}(x - q/2)^2 -  \frac{1}{2}(x + q/2)^2 - x^2} \\
&=   \sqrt{\pi/2} \, e^{-\frac{1}{4} q^2 - \frac{1}{2} (t_2 - t_1)^2 + (t_2 - t_1) q } \\
&=   \sqrt{\pi/2} \, e^{-\frac{1}{4} q^2} \sum_j H_j(q/\sqrt{2}) \frac{(t_2 - t_1)^j}{\sqrt{2}^j j!},
\end{align}
from which we obtain
\begin{align}
G_{mn}(q) &=  \frac{1}{\sqrt{\pi 2^{m+n} m! n! }} \partial_{t_1}^m  \partial_{t_2}^n O(0, 0) \\
&= A_{mn} \frac1{\sqrt{2}} e^{-\frac{1}{4} q^2}  H_{m+n}(q/\sqrt{2}).
\end{align}
Reinserting $\ell_B$ gives the summand appearing in Eq.~\eqref{eq:DLG}.

\subsection{Details of the magnetic Bloch basis} \label{app:MBB}

As shown in Fig.~\ref{fig:lattices}, we take for primitive translation vectors
\begin{align}
\mathbf{a}_1 &= a(0, 1), \\
\mathbf{a}_2 &= a(\sin\theta, \cos\theta),
\end{align}
where $a$ is the intervortex separation [related to the magnetic length via Eq.~\eqref{eq:theta}]. For ``$\pi$ flux'' per unit cell ($\frac{\phi}{\phi_0} = \frac{p}{q} = \frac12$), the corresponding magnetic translation operators are taken to act on the Landau gauge orbitals $\ket{k_y}$ (suppressing Landau-level index) as follows:
\begin{align}
T_1 \ket{k_y} &= e^{-i k_y a} \ket{k_y}, \\
T_2 \ket{k_y} &=
    \begin{cases}
        \ket{k_y + Q/2} & \mathrm{(square)} \\
        \ket{k_y + Q/2} e^{-i\pi(k_y/Q + 1/4)} & \mathrm{(triangular)}
    \end{cases}, \label{eq:T2action}
\end{align}
where recall $Q \equiv \frac{2 \pi}{a}$. $T_1$ and $T_2$ do not commute but rather satisfy the magnetic algebra $T_1 T_2 = -T_2 T_1$; thus, the \emph{magnetic Bloch basis}---as constructed in Eq.~\eqref{eq:MagBlochBasis}---is that which simultaneously diagonalizes $T_1$ and $T_2^2$. To obtain the second line of Eq.~\eqref{eq:MagBlochBasis}, we have used
\begin{align}
(T_2)^{2r} \ket{k_y} &=
    \begin{cases}
        \ket{k_y + r Q} & \mathrm{(square)} \\
        \ket{k_y + r Q} e^{-i\pi(2 r k_y/Q + r^2)} & \mathrm{(triangular)}
    \end{cases}, \label{eq:T22raction}
\end{align}
which along with the relation $Q \ell_B^2 = 2 a \sin\theta$ allows both lattice types to be treated at once. Note that the additional $k_y$-independent phase for the triangular lattice case in Eq.~\eqref{eq:T2action} is chosen so as to leave the resulting phase in Eq.~\eqref{eq:T22raction} void of terms proportional to $r$.

\subsection{Evaluating magnetic Bloch basis pairing matrix elements $\Delta_{mn}(\kvec)$ for arbitary $m,n$ [Eq.~\eqref{eq:Dmn}]} \label{app:DeltaDerivatives}

The full expression analogous to Eq.~\eqref{eq:D00_Bloch} for arbitrary $m,n$ is
\begin{widetext}
\begin{align}
\Delta_{mn}(\mathbf{k}) = \Delta_0 \, A_{mn} \sum_r e^{-i k_x (2k_y + r Q)\ell_B^2 - \frac{1}{4}(2k_y + r Q)^2 \ell_B^2 + i\pi\cos\theta\,r^2} H_{m+n}\left[\frac{(2 k_y + r Q) \ell_B}{\sqrt{2}}\right].
\end{align}
To evaluate this, we delay dealing with the Hermite polynomial by using the simple identity, valid for any $n$th order polynomial, $H_n(x) = H_n(\partial_{t}) e^{t x}|_{t = 0}$:
\begin{align}
H_{m+n}\left[\frac{(2 k_y + r Q) \ell_B}{\sqrt{2}}\right] &= H_{m+n}(\partial_{t})e^{t(2 k_y + r Q) \ell_B / \sqrt{2}} \\
&= H_{m+n}\left(\frac{i \partial_t}{\ell_B \sqrt{2}}\right) e^{-i t (2k_y + r Q) \ell_B^2}.
\end{align}
Equation~\eqref{eq:Dmn} can then be obtained after a few lines of algebra:
\begin{align}
\Delta_{mn}(\mathbf{k}) &= \Delta_0 \, A_{mn} H_{m+n}\left(\frac{i \partial_t}{\ell_B \sqrt{2}}\right) \sum_r e^{-i k_x (2k_y + r Q)\ell_B^2 - \frac{1}{4}(2k_y + r Q)^2 \ell_B^2 + i\pi\cos\theta\,r^2} e^{-i t (2k_y + r Q) \ell_B^2} \\
&= \Delta_0 \, A_{mn} H_{m+n}\left(\frac{i \partial_t}{\ell_B \sqrt{2}}\right) \sum_r e^{-i (k_x + t) (2k_y + r Q)\ell_B^2 - \frac{1}{4}(2k_y + r Q)^2 \ell_B^2 + i\pi\cos\theta\,r^2} \\
&= A_{mn} H_{m+n}\left(\frac{i \partial_{k_x}}{\ell_B \sqrt{2}}\right) \Delta_{00}(\mathbf{k}),
\end{align}
where in the last line we have used the expression for $\Delta_{00}(\kvec)$ in Eq.~\eqref{eq:D00_Bloch}. We can thus obtain arbitrary $\Delta_{mn}(\mathbf{k})$ by simply taking appropriate derivatives of $\Delta_{00}(\mathbf{k})$.
\end{widetext}

\bibliography{LLSC}

\end{document}